\documentclass[aps,showpacs,prb,twocolumn,supperscriptaddress]{revtex4-1}
\usepackage{times,xspace}
\usepackage{amsbsy,amssymb,amsmath,bm}
\usepackage{graphicx,color,epsfig,rotate}
\begin{document}

\title{Scaled free energies, power-law potentials, strain pseudospins and quasi-universality for  first-order
structural transitions}
\author{S.R. Shenoy$^*$, T. Lookman$^\dag$, and A. Saxena$^\dag$}
\address{$^*$School of Physics, University of Hyderabad, Hyderabad 500046, India, \\
$^*$International Centre for Theoretical Physics, Trieste
34014, Italy\\ $^\dag$Theoretical Division, Los Alamos National
Laboratory, Los Alamos, NM 87545, USA}

\date{\today}


\begin{abstract} We consider ferroelastic first-order phase transitions with $N_{OP}$ order-parameter strains entering  Landau free energies  as  invariant polynomials, that have   $N_V$ structural-variant Landau minima. The total free energy includes (seemingly innocuous) harmonic terms, in the $n = 6 -N_{OP}$ {\it non}-order-parameter strains. Four 3D  transitions are   considered,  tetragonal/orthorhombic, cubic/tetragonal, cubic/trigonal and cubic/orthorhombic unit-cell distortions, with  respectively,  $N_{OP} = 1, 2, 3 $ and  $2$; and $N_V = 2, 3, 4$ and $6$. Five 2D transitions are also considered, as simpler examples. Following Barsch and Krumhansl, we scale the free energy to absorb most material-dependent elastic coefficients into an overall  prefactor, by scaling in
 an overall elastic energy density; a dimensionless temperature variable; and the spontaneous-strain magnitude at transition   $\lambda <<1$.
  To leading order in $\lambda$ the scaled Landau minima become material-independent, in a kind of 'quasi-universality'. The scaled   minima  in $N_{OP}$-dimensional order-parameter space, fall at the centre and at the $N_V$ corners, of a
transition-specific polyhedron inscribed in a  sphere, whose radius is unity at transition.  The `polyhedra'  for  the four 3D transitions are respectively, a line, a triangle, a tetrahedron, and a hexagon. We minimize the  $n$ terms harmonic in the  non-order-parameter strains,  by substituting solutions of the 'no dislocation'  St Venant compatibility constraints,  and explicitly obtain  powerlaw anisotropic, order-parameter interactions, for all transitions. In a  reduced discrete-variable description,  the competing minima of the Landau  free energies 
 induce unit-magnitude  pseudospin vectors, with $N_V +1$ values, pointing  to the polyhedra corners and the (zero-value) center. The total scaled free energies then become                 $\mathbb{Z}_{N_V + 1}$ clock-like pseudospin hamiltonians, with  temperature-dependent local Landau terms,  nearest-neighbor Ginzburg couplings, and powerlaw St Venant interactions that drive the elastic domain-wall texturing. The  scaled free energies can be used in relaxational or underdamped dynamic simulations, to study  ferroelastic  strain textures and their dynamical evolution pathways. The pseudospin models  can similarly be studied via  local meanfield treatments, and  Monte Carlo simulations. 
 \end{abstract}
\maketitle

\vskip 0.5truecm


\section{Introduction}
Although first order phase transitions predominate in nature, second-order transitions have attracted much theoretical attention, because
of the taxonomically simplifying concept of `universality classes'
\cite{R1}.  Diverse materials in the same universality class have
common critical exponent values  as a scaled temperature $ T/T_c \rightarrow 1$, that depend only on the
dimensionality $d$ of coordinate space; the  number $N_{OP}$ of order parameter 
components; and the number $N_V$ of degenerate energy minima or
`variants'. Spin models are prototypical, with $d=2$, $N_{OP}=1$,
$N_V=2$ for a 2D Ising model. The symmetry-breaking
transition is signalled  by spontaneous nonzero averages of a spin
component, as one of the degenerate-minima variants in  order parameter
({\it OP}) space is picked out. First-order transitions by contrast, seem to be
inherently material specific, as they lack a divergent length scale
to induce universality, by rendering irrelevant the finite-scale
material parameters.  On the other hand, precisely because critical
fluctuations are unimportant, an approach focusing on
free energy minima is more reliable. If the free
energies could be scaled to make at least these minima independent of the material coefficients, one would have a kind of `quasi-universality' for first-order transitions.

Ferroelastic, displacive structural transformations
\cite{R2,R3,R4,R5,R6}  as in the austenite/martensite transition are (mostly) first order, symmetry lowering
transitions, with the  discrete symmetries of high/low temperature lattices often having a
 group/subgroup relationship.   Lattices in $d$ dimensions have
$d(d+1)/2$  Cartesian strain-tensor components, whose linear combinations are  the same number of physical strains,  of which $N_{OP}$ are the order parameters.   The Landau variational free energies are sums of high-order polynomial
invariants in the OP, with many material-specific anharmonic elastic coefficients \cite{R6,R7},  that take effort to extract from experiments. With a single high temperature zero-strain state,
and $N_V$ other structural variants apppearing as  temperature is lowered, 
there are $N_V +1$ degenerate Landau minima  at the first-order transition
temperature. Twinning, or spatial coexistence of competing structures separated by oriented domain walls, is common \cite{R2,R8}.
The orientation comes from long-range elastic forces \cite{R8} or powerlaw anisotropic interactions \cite{R11,R12,R13,R14,R15,R16,R17} between the $N_{OP}$ order parameters, that are present  but hidden, in the conventional displacement representation. As will be seen, these    effective interactions \cite{R11}, arise from  a {\it constrained} minimization of  free energy  harmonic terms in  the $n = d(d+1)/2  - N_{OP}$  non-order-parameter strains,  subject to  St. Venant compatibility conditions \cite{R9,R10}.

Barsch and Krumhansl (BK) \cite{R18} have scaled the  Landau free energy of a 2D square/rectangle transition, and the 3D cubic/tetragonal transition using three scaling parameters,
to a dimensionless form that is internally independent of elastic coefficients. A conceptually important dimensionless temperature variable
$\tau(T) \equiv (T - T_c) / (T_0 - T_c) $  can be introduced, that absorbs elastic coefficients of the quadratic term in the OP-strain magnitude. It is unity $\tau(T_0)=1$ at the first order transition temperature $T=T_0$; while it
 vanishes $\tau(T_c) = 0$ at the lower spinodal $T_c$.
 
 In this paper, we generalize this BK procedure for the Landau free energies of  four  3D and five 2D transitions, that can have  $N_{OP}=1,2,3$ order parameters, and $N_V = 2,3,4,6$ variants, to absorb the (often unknown)  material coefficients \cite{R6,R7}, into an overall energy prefactor. The four $3D$ transitions are tetragonal to orthorhombic, cubic to tetragonal, cubic  to trigonal; and cubic to orthorhombic, with their number of unscaled material coefficients respectively $N_{mat} =3,3,4$ and $6$.  The 3D transitions we have chosen are relevant for functional materials
: tetragonal/ orthorhombic (as for high-temperature superconductors
 such as YBCO); cubic/ tetragonal (as for shape memory alloys such as FePd); cubic/orthorhombic (as for ferroelectrics such as $BaTiO$); and cubic/trigonal (as for  CMR oxides such as $LaSrMnO$).
We show that,  apart from the overall energy-density coefficient, the scaled free energies evaluated at the $N_V +1$ minima are material-independent in the first three 3D cases, and weakly dependent on a single material parameter in the cubic-orthorhombic case. The minima fall  at the center and corners of a transition-specific polyhedron inscribed in a sphere, of unit radius at transition. The `polyhedron' for  $N_{OP} = 1,2$ or $3$ can be a line, a triangle, a hexagon or a tetrahedron. We evaluate the St Venant compatibility potentials through their $N_{OP} \times N_{OP}$ matrix kernels, for all the nine transitions using a  constraint substitution method, that is more direct  than a Lagrange multiplier \cite{R11} method. The scaled total  free energies can be used in over-damped, or under-damped OP strain dynamics (that includes Langevin noise terms with powerlaw anisotropic  spatial correlations) \cite{R12,R13}. In a reduced, discrete-variable description that retains only the Landau minima,  the total scaled free energies induce clock \cite{R19} like `$\mathbb{Z}_{N_V + 1}$' hamiltonians  in terms of  unit-magnitude, $N_{OP}$-dimensional vectors, pointing to $N_V + 1$ values. The    hamiltonians are  bilinear in the pseudospins,  with (temperature-dependent)
  quadratic on-site contributions  from the Landau term; 
nearest-neighbor ferromagnetic  couplings from the Ginzburg term; and powerlaw anisotropic  interaction potentials 
 from the St Venant term. 
 The pseudospin hamiltonians can be used for  local meanfield treatments, and Monte Carlo simulations, of ferroelastic textures. The models are  also be relevant for complex functional materials, with lattice strains coupled to intracell charge, spin and orbitals \cite{R17}.
  
  In more detail,  the transition-specific scaling procedure  involves a choice of three scaling parameters: a
typical spontaneous strain magnitude $\lambda$ that is small; a typical elastic energy density $E_0$; and a (Landau) first-order
transition temperature $T_0$ chosen  such that the scaled BK
temperature variable $\tau(T_0)=1$,  and the scaled OP magnitude is unity at transition. The smallness of  
$\lambda << 1$ justifies a finite-sum truncation of the expansion in invariants; and a neglect of geometric nonlinearities inside
 the Lagrangian strains, as a perturbative first approximation.  Working to leading order in $\lambda$,  yields  a  `quasi-universal' scaled Landau term that has an overall, material-dependent energy density $E_0$, as mentioned. The total free energy has Ginzburg and St Venant terms that determine elastic  domain-wall texturing, and  carry (two) material-dependent coefficients. Elastic pseudospin models in the context of martensites have been considered by several groups \cite{R20,R21,R22,R23,R24,R25}, but the models obtained here are explicitly induced by  scaled free energies.
 
In Sec. II we set up the general  BK scaling procedure for 2D and 3D  free energies. We then consider transitions in  increasing number of variants, $N_V = 2, 3, 4, 6$. In Sec. III we consider two-variant, single order parameter  $N_V =2, N_{OP}=1$ cases,
namely  the 3D
tetragonal/orthorhombic transition; and  in 2D  the square/rectangle or square/ rhombus;  and rectangle/oblique-polygon transitions. Section IV considers cases 
with $N_V =3$ and $N_{OP}= 2$, namely  the 3D cubic/ tetragonal, and 2D triangle/ centered-
rectangle transitions.
Section V considers $N_V = 4$ cases:  the 3D  cubic/ trigonal ($N_{OP}=3$), and 2D square/ oblique ($N_{OP} =2$). 
Section VI presents the $N_V = 6, N_{OP} =2$ case of  3D cubic/ orthorhombic and 2D triangle/ oblique  transitions.  Section VII
 obtains the reduced pseudospin hamiltonians, for all the transitions. Section VIII outlines possible simulation approaches, while the last Section  IX is a summary. An overview in Table I collects the generic numbers of the scaled Landau free energies, common to all materials with a given transition in the same quasi-universality class. The Appendix derives compatibility kernels for the four 3D and five 2D transitions considered, through the direct substitution method, contributing to a 'library of kernels' for use in simulations. 
 
 \section{SCALING PROCEDURE}

 Here  (A) we define elastic variables, and  (B) state the general scaling procedure. 
\subsection{  Distortion and strains}

  The distortion tensor \cite{R9}  $ D_{\mu \nu}$ can be defined in terms of gradients of the displacement vector $\vec u (\vec r)$ of points in a continuum medium, by

$$D_{\mu \nu} (\vec r) =  \partial u_{\mu}(\vec r) /\partial r_{\nu} ,~~(2.1a)$$

\noindent where $\mu , \nu$ run over $x, y, z$; or  in Fourier space, 

$$D_{\mu \nu} (\vec k) = i  u_{\mu}(\vec k) k_{\nu}.~~~~~(2.1b)$$ 
 The  ${\bf D}$ tensor is a sum ${\bf D} \equiv {\bf e} + {\bf w}$ of a symmetrized ($\bf e$) and antisymmetrized ($ \bf  w$) distortion (or local rotation) tensor. With  'T'  a transpose,

$${\bf e}(\vec r) \equiv \frac{1}{2} [ {\bf D} + {\bf D}^T] ; ~~{\bf w}(\vec r) \equiv \frac{1}{2} [ {\bf D} - {\bf D}^T].~~(2.2) $$  
 We will refer to the symmetrized distortion tensor $e_{\mu \nu}$ as the `strain' tensor,  and take it as the working variable, in a {\it strain representation}. It is distinct from the 'Lagrangian-strain' $E_{\mu \nu}$, that is a derived quantity \cite{R5} defined below. 
  
  In the nonuniform case, the  six symmetrized distortions $e_{\mu \nu} (\vec r)$  cannot vary
arbitrarily, if lattice integrity is to be maintained (i.e. no defects such as
dislocations).  They are linked by
the St. Venant's 'compatibility' equations \cite{R9,R10} of 1864 that ensure distorted unit cells fit together in a smoothly compatible way. With a
double-Curl operation defining an Incompatibility or `Inc' operation, in coordinate and Fourier space, the  Cartesian-component constraints on  ${\bf e} (\vec r)$ are 

$$Inc[{\bf e} (\vec r)]\equiv \vec{\nabla} \times (\vec{\nabla} \times {\bf e} (\vec r))^T =0;~(2.3a)$$
$$Inc [{\bf e}(\vec k)] \equiv \vec k \times {\bf e}(\vec k) \times \vec k =0.~~(2.3b)$$
(Similar equations hold for ${\bf w}(\vec r)$, but  for the transitions considered here we take spontaneous local rotations to be zero.)

From (2.1),  in the {\it displacement representation} where $\vec u (\vec r)$ is the working variable, compatibility is satisfied as an identity since $Curl (Grad) \equiv 0$.  Baus and Lovett  \cite{R10} have proposed  the distortion
 tensor $ {\bf D}$ be taken as the working variable, 
 with the St Venant compatibility equations
then being field equation constraints to be satisfied, rather than identities.  In an electromagnetic analogy \cite{R10} this is like working with the magnetic induction $\vec B$ rather than the vector potential $\vec A$,  with $Div \vec B = 0$ then a Maxwell field equation to be satisfied, rather than a $Div (Curl) \equiv 0$ identity.   The change to a distortion-tensor variable is natural and useful, since the free energy depends directly on the distortion; and the constrained minimization in the distortion
reveals powerlaw anisotropic potentials \cite{R11,R12,R13,R14,R15,R16,R17}, that are hidden in the more conventional displacement representation.

The elastic free energy is invariant under global uniform rotations \cite{R5}. Consider a line-element in the elastic medium,  described by a small imbedded column-vector $\vec a$ at a site $\vec r$. Under distortion of the medium, it locally stretches or rotates to a vector ${\vec A}(\vec r)= [ {\bf 1} +
{\bf D}(\vec r) ]\vec{a}$. Free energies
 $F$  can depend only on the scalar product ${{\vec A}(\vec r)^T}.{\vec A}(\vec r) = {\vec a}^T .{[ {\bf 1} + {\bf D}]^T} [{\bf 1}+ {\bf D}].{\vec a} \equiv {\vec a}^T .[ {\bf 1} + 2 {\bf E}].{\vec a}$, where  ${\vec A}^T, {\vec a}^T$ are row vectors. The free 
 energy $F(E_{\mu \nu} )$ thus depends on components of the `Lagrangian-strain' tensor \cite{R5} $E_{\mu \nu}$, that is a derived quantity, defined in terms of the basic distortion variable $D_{\mu \nu}$ by

$${\bf E}(\vec r) \equiv   \frac{1}{2} [ {\bf D}^T + {\bf D}] + \frac{1}{2}  {\bf D}^T {\bf D} = {\bf e} (\vec r) + {\bf g} (\vec r),~~(2.4)$$
 
\noindent where   the `geometric nonlinearity' is ${\bf g}  \equiv \frac{1}{2}  {\bf D}^T {\bf D}$.

   It is convenient to consider $d(d + 1)/2$ {\it physical} strains  $e_\alpha$ that are linear combinations of the $d(d + 1)/2$ Cartesian strains $e_{\mu \nu}$. For 2D lattices, with three Cartesian components, they
are $e_\alpha$ with $\alpha =1, 2, 3$: the compressional ($e_1$), 
   deviatoric ($e_2$), and shear ($e_3$) strains. With $X,Y$ transforming as 2D Cartesian coordinates, the physical strains transform as 
   $e_1 \sim X^2 +Y^2 , e_2 \sim X^2 - Y^2, e_3 \sim XY$,  and are defined as \cite{R13}

$$e_1 (\vec r) = \frac{c_1}{2} (e_{xx} + e_{yy});e_2 (\vec r)
  =\frac{c_2}{2} (e_{xx} - e_{yy});$$
  $$e_3 (\vec r) = \frac{c_3}{2} (e_{xy} +
  e_{yx})  = c_3  e_{xy}.~~(2.5a)$$
The physical Lagrangian-strains are similarly defined, so 
$$E_1 (\vec r) = \frac{c_1}{2} (E_{xx} + E_{yy}) = e_1 + g_1 ;$$
$$E_2 (\vec r) = \frac{c_2}{2} (E_{xx} -  E_{yy}) = e_2 + g_2; $$
$$E_3 (\vec r) =  c_3 E_{xy} = e_3  + g_3 ;~~~(2.5b)$$
where the physical geometric-nonlinearities are $g_1 = \frac{c_1}{2} (g_{xx} + g_{yy})$, $g_2 = \frac{c_2}{2} (g_{xx} - g_{yy})$, $g_3 =  c_3 g_{xy}$. For square unit-cells in the high-temperature phase, the normalizing coefficients are chosen as $c_1 = \sqrt{2} = c_2$, and $c_3 = 1$, while for triangular unit cells 
$c_1 = c_2 = c_3 = 1$.
  
  For 3D cubic lattices with six Cartesian strain components, the physical strains $e_\alpha$ with $\alpha =1,2,..6$ are the compressional ($e_1$), 
   deviatoric ($e_2 , e_3$), and shear ($e_4, e_5, e_6$) symmetrized-distortions. These transform as one, two, and three dimensional irreducible representations of the cubic point group. For the cubic lattice, with
    $X,Y,Z$ transforming as 3D Cartesian coordinates, the physical strains transform as $e_1 \sim X^2 +Y^2 +Z^2, e_2 \sim X^2 - Y^2, e_3 \sim  X^2 + Y^2 - 2Z^2 , e_4 \sim YZ, e_5 \sim ZX, e_6 \sim XY$, 
    and are defined as \cite{R14,R16}
   
$$e_1 (\vec r) = \frac{1}{\sqrt{3}} (e_{x x} + e_{y y}+e_{z z});$$ 
$$e_2 (\vec r) =\frac{1}{\sqrt{2}}(e_{x x} - e_{y y}); ~~  e_3 (\vec r) =\frac{1}{\sqrt{6}}( e_{x x} + e_{y y} -2 e_{zz})~;$$
$$e_4 ({\vec r})  =  2e_{y z}, e_5 ({\vec{ r}}) = 2e_{z x}, e_6 (\vec r) =  2e_{x y}.~~~~(2.6a)$$
The physical Lagrangian-strains $E_\alpha$  are similarly defined so  e.g. $E_1 = (E_{xx} + E_{yy} + E_{zz})/ {\sqrt{3}} = e_1 + g_1$ where
$g_1 \equiv (g_{xx} + g_{yy} + g_{zz})/{\sqrt{3}}$, and so on.   

For the tetragonal lattice \cite{R16},
$$e_1 (\vec r) = \frac{1}{\sqrt{2}} (e_{x x} + e_{y y});$$
$$e_2 (\vec r) =\frac{1}{\sqrt{2}}(e_{x x} - e_{y y}); ~ e_3 (\vec r) = e_{zz}. ~~~~(2.6b)$$
and the physical shears are the same as (2.6a).
These transform as irreducible representations of the tetragonal point group.
   
   As noted in the Appendix, the St Venant constraint (2.3b) in Fourier space can be written in terms of the physical strains as

$$\sum_{\alpha = 1,2...6}O^{(s)}_{\alpha} e_\alpha(\vec k)  =0,~~(2.7)$$

\noindent where  $O^{(s)}_{\alpha} (\vec k)$ are compatibility coefficients appropriate to the symmetry, labelled by shear components, $s =3$ in 2D (one constraint), and  $s = 4,5,6$ in 3D (three constraints). This is used in the Appendix to calculate the compatibility potentials.
 
     The physical strains can be separated into $N_{OP}$ order parameter strains labelled by $\alpha = \ell$ or $\{e_\ell\}$; and 
    $n = [\frac{1}{2} d ( d + 1) - N_{OP}]$ non-OP strains, labelled by $\alpha = i$ or  $\{e_i\}$. (An associated separation of the physical Lagrangian-strains for each transition into $\{E_\ell\}$ and $\{E_i\}$ is also made.) 
    For a ferroelastic transition, the free energy $F = F_L + F_{non} + F_G$ . The Landau part of the free energy $F_L (E_\ell)$ can be written as a sum of $p$-th order polynomial invariants
    in  the OP Lagrangian-strains, that can be found by direct evaluation \cite{R11,R12,R13,R14,R15} for simple cases.
  The non-OP free energy $F_{non} (E_i)$ is taken as
   harmonic in the 
   non-OP Lagrangian-strains. For simplicity we neglect symmetry-allowed anisotropic gradients \cite{R14,R15}, and consider only  OP gradient-squared costs  in the Ginzburg term $F_G({\vec \nabla}E_\ell)$. For all 2D transitions, an exhaustive evaluation of the OP and OP-gradient invariants, and allowed OP/non-OP couplings, was carried out  through the program ISOTROPY of Stokes and Hatch \cite{R15}.

   Then with a sum or integral ($\sum_{\vec r} \rightarrow \int d^d r / a_0 $) over all positions $\vec r$ where $a_0$ is a lattice scale, the total variational free energy in terms of free energy densities $f$ is

$$F=\sum_{\vec r} f_{L} (E_\ell)+ f_{G}({\vec{\nabla}}E_\ell)+f_{non}(E_i).~~(2.8)$$

\noindent The Landau term  $f_L$ is not just an arbitrary Taylor series, but is a finite-sum expansion in terms of symmetry-allowed, invariant  polynomials of physical OP Lagrangian-strains

$$f_L = C^{(2)} (T) I_2 + \sum_{p =3,...p_{max}} \sigma_p C^{(p)} I_p(E_\ell). ~~(2.9a)$$
The second order invariant
$I_2=\sum_{\ell} E_{\ell}^2$ is common to all transitions, and is separated out. The anharmonic elastic coefficients $ C^{(p)} > 0$ are temperature-independent, and $\sigma_p = +1, -1$ are chosen signs, to get $N_V$ minima. We consider an elastic coefficient
$C^{(2)} (T)= (T-T_c) {C_0}^{(2)}$ that  would partially soften to zero on cooling to a
temperature $T=T_c$, but is preempted by the first order transition.  The Ginzburg and non-OP terms, with domain wall cost parameter $b$ and $A^{(i)}$ 
elastic coefficients \cite{R26}, are

$$f_G = \sum_{ \ell} b (\vec{\nabla}E_{\ell} (\vec r))^2;~ f_{non}=\sum_{i}\frac{1}{2} A^{(i)} {{E_i}^2} (\vec r).  ~(2.9b)$$
The temperature-dependence of sound velocities \cite{R27} experimentally determine the linear slope ${C_0}^{(2)}$, and the extrapolated temperature $T_c$. The curvature of phonon spectrum at long wavelengths \cite{R27} determine $b$, that is related to domain-wall energy costs at short wavelengths. We take it to be positive $b > 0$;  but for $b < 0$,
one can add phenomenologically  a symmetry allowed positive-coefficient, fourth-order gradient term, for stability.

\subsection{ General scaling and minimization procedure}

 We now follow a Barsch-Krumhansl procedure \cite{R18}: scaling all Cartesian distortions $D_{\mu \nu} \rightarrow \lambda D_{\mu \nu}$ or $e_{\mu \nu} \rightarrow \lambda e_{\mu \nu}, w_{\mu \nu} \rightarrow \lambda w_{\mu \nu}$ in a typical value $\lambda$; scaling  all
 energy terms in $E_0$ ; and defining a transition temperature $T_0$. These three parameters are chosen  in terms of the material-specific elastic constants, to make the Landau free energy simple. The parameters have physical meanings:  since scaled terms are of order unity, the overall prefactor $E_0$ is the elastic energy per unit cell;  the temperature $T_0 (> T_c)$ is the first-order transition temperature that pre-empts the second-order  elastic-constant softening  at $T_c$; and $\lambda$ is  the  spontaneous-strain magnitude at $T_0$, since the scaled strain is chosen to be unity at transition.
 
 Since the physical  distortions are linear combinations of the Cartesian distortions, they change as $e_\alpha \rightarrow \lambda e_\alpha$,  and the physical Lagrangian-strains change as 

$$E_\alpha \rightarrow \lambda E_\alpha (\lambda) \equiv \lambda (e_\alpha + \lambda g_\alpha ).~~~~~(2.10)$$ 
Henceforth $e_\alpha$ is the {\it scaled} symmetrized-distortion or scaled strain. The (scaled) geometric nonlinearity $g_\alpha$ carries a prefactor         $\lambda$. The free energy changes as

$$F(E_\alpha) \rightarrow F[\lambda E_\alpha (\lambda)] \equiv E_0 \bar{F} [E_\alpha (\lambda)], ~~~(2.11a)  $$ 

\noindent defining a  dimensionless  $\bar{F}[ E_\alpha (\lambda)] =\sum_{\vec r} \bar{f} [E_\alpha (\lambda)]$ 
where  the scaled dimensionless free-energy densities are 
$$\bar{f}[E_\alpha (\lambda)] \equiv \bar{f}_L[E_{\ell} (\lambda)] + \bar{f}_G [{\vec \nabla} E_{\ell} (\lambda)] +\bar{f}_{non}[E_i (\lambda)].~(2.11b)$$
These terms contain  dimensionless
 scaled  coefficients that absorb the `external'  $\lambda$ powers of (2.10) as

$$C_p \equiv C^{(p)} \lambda^p /E_0 ; ~A_i \equiv A^{(i)} \lambda^2/E_0;~\xi^2 \equiv b \lambda^2/{a_0}^2 E_0 .~(2.12)$$
Here a dimensionless  length $\xi$ has a lattice constant $a_0$ scale, from  a substitution as below in the Ginzburg term $\vec{\nabla} \rightarrow \vec{\Delta} / a_0$, where $\vec{\Delta}$ is a discrete-difference operator on a computational grid.

A scaled temperature \cite{R18} $\tau$ absorbs the harmonic-term material dependence, and is defined as  

$$ \tau (T) = (T- T_c){C_0}^{(2)}  \lambda^2 /E_0 \equiv (T-
T_c)/(T_0 - T_c), ~~~(2.13) $$

\noindent where the transition temperature $T_0 > T_c$ is determined by requiring $\tau(T_0) = (T_0 - T_c){C_0}^{(2)} \lambda^2  / E_0 =1$ or

$$T_0 = T_c + E_0 /({C_0}^{(2)} \lambda^2 ) .~~~~~(2.14)$$

At  the first-order Landau transition temperature
  at  a universal value $\tau(T_0) =1$, the nontrivial or `martensite' wells are degenerate with the trivial  or `austenite' well; while $T_c$ is the lower spinodal, where at a universal value $\tau(T_c) =0$, the metastable
austenite well disappears.
The upper spinodal $T_{up}$ where the martensite-variant  wells disappear, turns out also to have a universal value
 $\tau (T_{up}) = \tau_{up}$ for  three of the four 3D transitions. 

A constrained minimization of the  scaled free energy $ \bar{f}[E_\alpha (\lambda)]$  of (2.11b) with respect to   $e_\alpha$,  
 would, for general $\lambda$, fix the parameters $\lambda, E_0$, and yield
an effective OP-OP interaction potential.
However because of geometric nonlinearities the calculation is involved, and we perturbatively evaluate the scaling parameters and compatibility potentials, as below. Corrections can in principle, be calculated.

    We make a simplifying assumption that the typical  spontaneous distortion
  is small compared to unity

  $$\lambda \ll 1,~~~~~~~~~(2.15)$$
and for most materials this is indeed a few per cent, $\lambda \sim 10^{-2}$.
Then the scaled physical Lagrangian-strains $E_{\alpha} (\lambda) = e_\alpha + \lambda g_\alpha$  are approximated by the scaled physical symmetrized-distortions,

$$E_{\ell} (\lambda) \simeq E_{\ell} (0)= e_{\ell};  E_{i} (\lambda) \simeq  E_{i} (0)=e_i  .~(2.16)$$
In fact, this approximation of dropping geometric nonlinearities in the Lagrangian-strains  is commonly made without specific comment.  In the conventional (unscaled) displacement representation it is implicitly justified as a long-wavelength truncation:  in Fourier space the strain tensor  is $\sim [{\vec q} {\vec u} (q)]_{\mu \nu}$, while the geometric nonlinearity  is $g_{\mu \nu} (\vec q) \sim {[{\vec q} {\vec u} (\vec q)]^2}_{\mu \nu}$, that is  higher order in $\vec q \rightarrow 0$. Instead, in the (scaled)  strain representation, the neglect of geometric nonlinearities is seen as the leading term in a small-parameter expansion in $\lambda$, that could be systematically corrected. 
  
  The scaled free energy densities  of (2.11b) then become $\bar{f} [E _\alpha (\lambda)] \simeq \bar{f} (e _\alpha)$, 
 to leading order in $\lambda$, separately for each distinct symmetry-invariant term,   
$$\bar{f} (e _\alpha) =\bar{ f_{L}} (e_\ell)  +
\bar{f_{G}}({\vec \Delta}e_\ell)+\bar{f}_{non}(e_i),(2.17)$$

\noindent From (2.9a), (2.9b), (2.12) and (2.13),

$$\bar{ f}_L(e_\ell)=  (\tau - 1) \sum_{\ell} {e_{\ell}}^2 + f_0(e_\ell),~~~(2.18a)$$

$$f_0(e_\ell) \equiv  \sum_\ell {e_\ell}^2 + \sum_p \sigma_p C_p I_p (e_{\ell}),~~~(2.18b)$$

$$\bar{f}_G({\vec \nabla}e_\ell) =  \sum_\ell \xi^2 (\vec{\Delta} e_{\ell})^2,~~
\bar{f}_{non}(e_i)  = \sum_{i}\frac{A_i}{2} {e_i}^2. ~(2.18c)$$
Here as mentioned previously, $\vec{\Delta} = (\Delta_x , \Delta_y, \Delta_z )$ has discrete forward-difference operator components on a cubic computational grid.

We pause to relate the unscaled harmonic coefficients  $A^{(i)}$ of (2.9) to the material elastic constants. The elastic energy is \cite{R26} 
$E = \frac{1}{2} \sum_{\alpha, \beta} C_{\alpha \beta} x_\alpha x_\beta$ where  with $\alpha = 1,2, ..6$,  the $\{ x_\alpha\}$  are Cartesian strains written as a column vector such as     $x_1 = e_{xx}, x_4 = 2 e_{yz}$.    In the cubic case, there are three independent elastic constants in the Voigt notation, $C_{11}, C_{12}, C_{44} (= C_{55}= C_{66})$. Writing Cartesian in terms of physical strains, through (2.6a) and (A9), the energy $E$ is diagonalized, and a comparison with (2.9) yields $C_{11} - C_{12} = C^{(2)} (T), C_{11} + 2C_{12} = A^{(1)}, C_{44} = A^{(4)}$. It is useful to introduce the  elastic anisotropy parameter  \cite{R26}  $A(T) \equiv  2 C_{44} / (C_{11} -C_{12})$, where  $A > 1$ (or $A < 1$) corresponds to greater stiffness in the body diagonal $<111>$ directions (or cubic axis $<100>$ directions).  Strongly anisotropic materials can have $A \sim 10$.  Then  from (2.12),  the scaled shear coefficient $A_4 (=A_5 = A_6)$, and the scaled compression coefficient $A_1$ are both in terms of the  elastic anisotropy parameter  $A(T=T_0)$ at transition,   

$$A_4 = A(T_0); ~~ A_1 = \gamma A(T_0);~~\gamma \equiv   [(C_{11} + 2 C_{12})/ C_{44}] . ~~(2.18d)$$
The elastic constant  ratio $\gamma = A_1 /A_4 (= A_1/A_5 =A_1/A_6)$, that enters  the compatibility potentials, can for simplicity  be set in simulations to a constant,  say  $ A^{(1)} / 2A^{(4)}  \sim 1$ as in \cite{R11} FePd. [For $x y$ plane distortions, the shear term is $(A_6/2) 4 {e_{xy}}^2$, so from (2.5a) the scaled 2D shear coefficient appearing later is also proportional to the anisotropy at transition.] Finally we note that from (2.12), the scaled Ginzburg coefficient is similarly $\xi^2 \sim (b/ a_0^2) A(T_0)$.

We now return to the main argument. A constrained minimization of the harmonic non-OP terms in (2.18c)  as in the Appendix yields the non-OP  in terms of the OP strains,   $e_i (\vec k) = \sum_{\ell} B_{i \ell} (\vec k) e_\ell (\vec k)$ where the $B_{i \ell}$ coefficients are  in terms of the
coefficients ${O_\alpha}^{(s)}$ of (2.7). Substituting back in the harmonic term, 
$ \bar{F}_{non}[e_i(e_{\ell})] \equiv \bar{F}_{compat}(e_{\ell})$ induces  the St Venant term 

$$\bar{F}_{compat}(e_{\ell})=\frac{1}{2} A_1\sum_{{\vec k},\ell,\ell'} U_{\ell\ell'}(\vec{k}) e_{\ell}(\vec{k})e_{\ell'}^\ast(\vec{k}). ~(2.19)$$
The compatibility kernel $A_1 U_{\ell \ell'}(\vec{k})\equiv\sum_i A_i B_{i\ell}(\vec{k}) B^\ast_{i\ell'}(\vec{k})$ is evaluated for each transition in the Appendix, and is essentially dependent only on the wave-vector direction $\hat k$, independent of the magnitude $|\vec k|$. In coordinate space, the compatibility potential
is hence an anisotropic powerlaw  with a falloff exponent equal to the dimensionality \cite{R11,R12,R13,R14,R15,R16,R17,R23} $U_{\ell \ell'}(\vec{R}) \sim 1/R^d$.  (Write the Fourier integral of $U_{\ell\ell'}(\hat {k})$and change the  wave-vector integration variable $|\vec k| \rightarrow |\vec k|/R$: the exponent simply comes from the phase space dimension.)

  We focus on  the Landau term $\bar{f}_L$. To find the minima in OP space 
it is convenient to work in polar coordinates, following Toledano and Toledano \cite{R6}. For example the $N_{OP}=2$ dimensional `vector'  in OP space is $\vec{e} = (\varepsilon \cos \phi ,\varepsilon \sin \phi )$,
  where the `radial' variable is

$$\varepsilon \equiv |{\vec{e}}| = [{\sum_{\ell} {e_{\ell}}^2} ]^{1/2},~(2.20)$$
 and  the Landau free energy density  is 

$$\bar{ f_L} (\vec e) =\bar{f_L} (\varepsilon, \phi) =(\tau-1) \varepsilon^2 + f_0 ( \varepsilon, \phi), ~(2.21)$$
where the transition-specific $f_0$ is temperature independent.
We demand that the nontrivial Landau minima are at  $m = 1,2,..N_V$ equivalent points $\{\varepsilon_m  ,\phi_m\}$,  with  the same radii $\varepsilon_m = \bar{\varepsilon}$ in OP space. The conditions are:

$$\frac{\partial \bar{f_L} (\varepsilon_m, \phi_m)}{\partial \varepsilon_m} = 2(\tau -1)
  \varepsilon_m + \frac{\partial f_0 (\varepsilon_m, \phi_m)}{\partial \varepsilon_m} =0,~~(2.22a) $$

\noindent locating the martensitic minima in the  radial direction at a temperature-dependent $\varepsilon =\varepsilon_m = \bar{\varepsilon} (\tau)$; and

$$\frac{\partial \bar{f_L} (\varepsilon_m, \phi_m )}{\partial
  \phi_m} =  \frac{\partial f_0 (\varepsilon_m, \phi_m)}{\partial
  \phi_m} =0,~~(2.22b)$$

\noindent locating the minima in the azimuthal direction at a temperature-independent  $\phi = \phi_m$.  At transition $\tau =1$, we also demand that the nonzero minima on a      $N_{OP}$-dimensional `sphere'  of
  radius $\bar{\varepsilon} (\tau =1)=1$, become
 degenerate with the trivial minimum $\bar{f}_L ({\vec e}=0) =0$. Hence

  $$f_0 (\varepsilon_m=1,\phi_m)=0.~~(2.22c) $$
Above an upper spinodal $\tau = \tau_{up}$ the radial solutions $\bar{\varepsilon} (\tau)$ become imaginary, and there is only the trivial austenite minimum. It is convenient  for later use to define the Landau free energy at minima

$$\bar{f_L}(\tau)  \equiv \bar{f_L} (\bar{\varepsilon}, \phi_m ) \equiv  {\bar{\varepsilon} (\tau)}^2 g_L (\tau), ~(2.22d)$$
where $g_L (\tau)$ changes sign at  the Landau transition.

 With $\tau$ defined,  we choose the remaining two scaling parameters $\lambda, E_0$ so the $f_0$ conditions  of (2.22b), (2.22c)  are
  satisfied.  It is useful from (2.21) to separate the angular dependence into a part $\Delta f_0 ( \varepsilon, \phi) \equiv f_0 ( \varepsilon, \phi)-  f_0 ( \varepsilon, \phi_m)$  that vanishes in the minimum angular directions,  so
  $$\bar{f_L} (\varepsilon, \phi) \equiv [(\tau-1) \varepsilon^2 + f_0 ( \varepsilon, \phi_m)] + \Delta f_0 ( \varepsilon, \phi),~(2.23a)$$
and  $\bar{\varepsilon} (\tau)$  is determined through minimization  of only 

$$\bar{f_L} (\varepsilon, \phi_m) = [(\tau-1) \varepsilon^2 + f_0 ( \varepsilon, \phi_m)].~ (2.23b)$$

There is always an overall material constant $E_0$ for the Landau energy $f_L = E_0 \bar{f_L}$, that  absorbs unknown higher-order elastic coefficients, and can be treated as a fitting parameter.  Since we work only to leading  order in $\lambda$, any  material-independence found in the scaled $\bar{f_L} ({\vec e})$  contributions  is  strictly speaking only {\it quasi}-universal.  Landau quasi-universality can be of three kinds:  (i) {\it strong}, i.e. the  scaled $\bar{f_L} (\varepsilon, \phi)$ is independent of material parameters for all $\varepsilon, \phi$;   (ii) {\it medium}, i.e.  material coefficients appear only in  $\Delta f_0$ that vanishes at $\phi = \phi_m$, so along minima angles,  $\bar{f_L} (\varepsilon, \phi_m)$ is material-independent; and finally  (iii) {\it weak}, with residual material-dependence in ${\bar{f_L}} (\varepsilon, \phi_m)$, and hence in the OP magnitude   $\bar{\varepsilon} (\tau)$ (that is however still unity at transition for all materials). The tetragonal/orthorhombic and cubic/ tetragonal;  the cubic/trigonal ; and the  cubic/orthorhombic  transitions  (with unscaled material coefficients of  respectively $ N_{mat} = 3, 3, 4$ and $6$), turn out  to have quasi-universality  in $\bar{f}_L$ of respectively the first,  second  and third kinds. Table I summarizes the generic numbers for all transitions considered, with different materials with the same transition falling into the same 'quasi-universality class'.

 Going back to  unscaled variables denoted by primes, the  unscaled entropy-density difference relative to the austenite  from the Landau term is 
  ${s_L} '(T) = -\partial f '_L({\bar{\varepsilon }} ' (\tau), \phi_m ) /\partial T$. Since the derivative of (2.22a) with respect to 
  ${\bar{\varepsilon}} ' (\tau)$ vanishes, only the explicit $\tau$-dependence of ${\bar f}_L$
contributes.  The scaled entropy-density difference  is

$$  \bar{s}_L (\tau) \equiv [(T_0 - T_c )/E_0 ] {s_L} ' ( \tau) =- {\bar{\varepsilon}(\tau)}^2, ~~(2.24)$$

\noindent  and is (minus) unity at transition.  Of course there are other free energy terms, 
and hysteresis from domain-wall textures, so this is just a formal result.We will consider {\it proper} ferroelastic transitions with free energy nonlinearities in the OP strain driving the transition (without intracell shuffles); with high/low temperature unit-cell symmetries having a group/subgroup relationship; and without coupling to other fields. (There are also improper ferroelastics, with  only harmonic  terms in strains, that are however  coupled to other fields such as electric polarization or magnetization, whose nonlinearities can induce a structural transition \cite{R3}.)

 In Secs. III--V we consider scaling of the Landau free energy (and
other terms), for four  3D transitions with $N_{OP} = 1,2,3$   and for five 2D transitions with $N_{OP} = 1,2$,  presented in increasing number of variants $N_V = 2,3,4,6$. Some cases had been scaled earlier \cite{R12,R13,R18} but are summarized here for completeness. The final results for the minima are
summarized in Table I.

\section{TRANSITIONS WITH  $N_V=2$}

We consider one 3D transition and two 2D transitions, all with $N_V=2$ low temperature
variants and a single order parameter (OP)  component
$N_{OP}=1$, with the number of non-OP variables $n = \frac{1}{2} d(d + 1)- N_{OP}$. The transitions are:  (a) tetragonal/orthorhombic in 3D $(n=5)$;  and (b) square/rectangle (that includes square/ rhombus); and (c) rectangle-oblique  cases (all $n =2$). See   Figs 1 and 2.

\subsection{Tetragonal/orthorhombic  case in 3D: $N_V = 2, N_{OP} = 1, n =5$}

 There are two deviatoric distortions $e_2 , e_3$ in 3D, and the single ($N_{OP} =1$) order parameter
is $e_2  \sim X^2 - Y^2$ that can change a tetragonal square-cross-section to an orthorhombic rectangular-cross-section. Since there are two possible such rectangular elongations
(along mutually perpendicular directions), one expects two possible variants, as in Fig 2.
The tetragonal point group G =$P4/mmm$ with $\nu_G = 16$ elements goes in a symmetry-lowering transition, to the orthorhombic subgroup g =$Pmmm$ with $\nu_g =8$ elements \cite{R6}. The orthorhombic group describes  symmetries of a unit cell with a particular rectangular orientation, corresponding to one variant. The ratio of the number of elements in a point group to that in a subgroup, or dimension of the coset $G/g$,  is  an integer, that we assume  \cite{R6} corresponds to the number of variants. Thus  here there are $N_V = \nu_G / \nu_g = 16/8=  2$ variants, as expected. The ratio of the number of (rotational) elements in the point group elements, has been taken to be the number of variants \cite{R2}, yielding the same result $8/4 =2$ here, and in other cases.  (See however,  Section VI below.)

For these two variants in a first
order transition, we need up to $p_{max} = 6$  even-order strain invariants $I_p$. The unscaled free energy, with sign choices $\sigma_4 = -1, \sigma_6 = +1$  is 
$f_L = C^{(2)} I_2 - C^{(4)} I_4 + C^{(6)} I_6$, where  $I_p = {e_2}^p$ are invariants, so there are $N_{mat} = 3$ material constants.  With  $e_2 \rightarrow \lambda e_2$  and $C_p \equiv \lambda^p C^{(p)} /E_0$ as in (2.12), the scaled Landau free energy density ${\bar f}_L = f_L / E_0$  is as in (2.21):

$${\bar f}_L (e_2) =  (\tau -1) e_2 ^2 +f_0  (e_2),~~~~~~(3.1)$$ 
where the temperature-independent $f_0$  is

$$\bar{f}_0 (e_2)= e_2^2 - C_4 e_2^4  + C_6 e_2^6 . ~~~~(3.2) $$

 The conditions  (2.22a), (2.22c)  for degenerate minima  are  $\partial
f_{0}/\partial\varepsilon + 2 (\tau - 1) \varepsilon =0$, and $ f_{0}(1)=0$. This fixes the two coefficients

$$C_4=2; ~C_6=1,~~~~~~(3.3)$$
achieved by choosing  the two scaling parameters as 

$$\lambda=(C^{(4)}/2C^{(6)})^{1/2}; ~~~E_0=C^{(6)} (C^{(4)}/ 2 C^{(6)})^3. ~(3.4)$$
Then $f_0$ becomes a perfect square, and

$$\bar{f}_L(e_2)=(\tau-1)e_2 ^2+ e_2 ^2(e_2 ^2 -1)^2 , ~~~~~(3.5)$$

\noindent manifestly showing the triple-minima degeneracy  at $f_L =0$ for $\tau =1$. The variant minima are at $\pm \bar{\varepsilon} (\tau)$ where the order parameter magnitude

$$\bar{\varepsilon} (\tau)=\left\{\frac{2}{3}[1+\sqrt{1-3\tau/4}]\right\}^{1/2},
~~~~~(3.6) $$

\noindent is unity at transition.  Here from (2.22d) the variable used later in pseudospin hamiltonians
is $\bar{f}_L / {\bar{\varepsilon}}^2 \equiv  g_L = \tau -1 + ({\bar{\varepsilon}}^2 - 1)^2$.
Barriers at ${\bar{\varepsilon}}_b (\tau)=\left\{\frac{2}{3}[1-\sqrt{1-3\tau/4}]\right\}^{1/2}$
 exist in the range  $0 < \tau< \tau_{up} = 4/3$. The barriers merge with the metastable martensite (or metastable austenite) minimum  at the upper spinodal $\tau =4/3$ (or lower spinodal
$\tau = 0)$. 

Since $\bar{f}_L [\bar{\varepsilon} (\tau)]$ is independent of material parameters, this is quasi-universality of the first kind, with
$f_L = E_0 \bar{f}_L$ having only an overall material dependence through $E_0$, that absorbs the  higher-order elastic constants.
  
 With the OP sign formally written as an angle, $e_2 = |\varepsilon| \cos \phi$ where $|e_2| = |\varepsilon|$, and the minima are at $\phi_m = 2(m - 1)\pi/ N_V$, with $m = 1,2 (=N_V)$,  where $\sin 2 \phi_m = 0$. 
 At transition, the nontrivial Landau minima fall at $N_V =2$ points
 at $\pm 1$ on a line in the  $N_{OP} = 1$ dimensional order parameter space, as in Fig 3 and Table I. The number of distinct martensite/martensite domain walls between variant pairs is $N_W =1$.

Had we included an eighth order invariant  $\sim C^{(8)} {e_2}^8$ in (3.2),    the minimum condition would become $\tau  -2 \varepsilon^2 + 3 \varepsilon^4 + 4 C_8 \varepsilon^6 = 0$. Here the new scaled coefficient is $C_8 =  C^{(8)} \lambda^8 /E_0 =( C^{(8)}/C^{(6)}) \lambda^2$. Assuming the ratio of eighth and sixth order unscaled constants is not too large,  the shift in the roots arising from $C_8 \sim \lambda^2 \ll 1$ is negligible.
 The extra eighth order invariant is thus 'irrelevant', in the sense that the polyhedral minima remain essentially unchanged, justifying the finite-sum restriction to $p_{max} = 6$ of the polynomial expansion. 

The scaled Ginzburg term  is  $\bar{f}_{G}=\xi^2({\vec \Delta} e_2)^2$. There are  $n=5$ non-OP strains, namely 
 the compression $e_1\sim X^2 + Y^2$;  the other deviatoric strain $e_3 \sim X^2 + Y^2 - 2 Z^2$;  and the three shears $e_4, e_5, e_6$.
 Using the three compatibility constraints of (2.7) to eliminate the shears, and minimizing in $e_1$,  the non-OP strains are determined by the OP. Substituting into  the harmonic non-OP terms yields an OP compatibility potential term as in (2.19):
 $\bar{f}_{non} (e_i) =  \sum_{ i = 1,3,4,5,6} ( A_i /2) {e_i}^2  \rightarrow \bar{f}_{compat} (e_2) $ where

$$\bar{f}_{compat} =  \frac{A_1}{2} U(\vec k) |e_2 (\vec k)|^2 ,~~~~~(3.7) $$
and the kernel $U(\vec k)$ is given in (A26) of the Appendix.

\subsection{Square/rectangle  case: $N_V =2, N_{OP} =1, n=2$}

The single 2D deviatoric distortion $e_2 \sim X^2 - Y^2$  turns a square to a rectangle. Since the rectangular elongation can be along two axes, one expects two variants, as in Fig 1. The point group  G = $p4mm$ for a square unit cell
has  $\nu_G = 4$ elements, while the subgroup g for the
rectangle is $p2mm$ with $\nu_g =2$ elements. Thus the number of
variants is \cite{R15} $N_V = \nu_G / \nu_g = 2$ as expected, and we again need up to $p_{max} = 6$  order invariants. The scaled Landau free energy $\bar{f}_L (e_2)$ is the same form as (3.5) above,  so
the minima are at the same $\pm \bar{\varepsilon} (\tau)$ of (3.6).
 
The scaled Ginzburg term is  $\bar{f}_{G}=\xi^2(\vec{\Delta} e_2)^2$. There are $n =2$ 
non-OP strains $e_1 \sim X^2 + Y^2, e_3 \sim XY$, and the harmonic free energy  term  is $ \bar{f}_{non}(e_i) = \frac{1}{2}  \sum_{i = 1,3} A_i |e_i (\vec k)|^2$. This induces a compatibility kernel \cite{R11,R12,R13,R15}  $U(\vec k)$ as in  (A4) of the Appendix.

The square/rhombus transition has $N_V = 2, N_{OP} =1, n=2$, and the 2D  shear physical distortion $e_3 \sim XY$ as the single order parameter.  However  this is not an independent transition \cite{R15}, since 
$e_2$ and $e_3$ interconvert through a global rotation of Cartesian  axes by $\pi /4$. Nonetheless, as an exercise the symmetry group
of the square is G = $p4mm$ with $\nu_G = 4$ components, while the rhombus
symmetry is g = P2 with $\nu_g =2$ elements, so $N_V =2$.  The scaled Landau
free energy is  $\bar{f}_L (e_3)$, formally as in (3.5), 

$$\bar{f}_L(e_3)=(\tau-1)e_3 ^2+ e_3 ^2(e_3 ^2 -1)^2 , (3.8) $$  
with two nonzero minima at the same values $e_3 = \pm \bar{\varepsilon} (\tau)$.

The Ginzburg term is $\bar{f_{G}}=
\xi^2(\vec{\Delta}e_3)^2$.  The $n=2$  non-OP strains are now 
compressional and deviatoric, $e_1\sim X^2 + Y^2$ and $e_2 \sim X^2 - Y^2$, with the harmonic \cite{R11,R12,R13}
$\bar{f}_{non} = \frac{1}{2} \sum_{i = 1,2} A_i  |e_i |^2$  inducing a kernel  $U(\vec k)$ as  in (A5) of the Appendix.

There is another symmetry-allowed  transition \cite{R13,R15},  namely the square/centered rectangle. In addition to strain, as in the square/rectangle, it  also involves a shuffle because of the center site, so is not considered here.

\subsection{Rectangle/oblique case: $N_V = 2, N_{OP} =1, n=2$}
 
 The shear physical distortion $e_3$ changes a rectangle to an oblique polygon, and is  the single $N_{OP} =1$ order parameter.
 The point group G =$p2mm$ with $\nu_G = 2$ elements goes to the subgroup g= $p2$ 
 with $\nu_g =1$ elements, so there are $N_V = 2$ variants. The
${\bar{f}}_L(e_3)$ Landau part is the same as the square/rhombus case of (3.8); however the $n=2$ non-OP
contributions are harmonic in the combinations $e_{\pm} = (e_1\pm e_2)/ 2$, as 
${\bar{f}}_{non} = \frac{A_{+}}{2} {e_{+} }^2 + \frac{A_{-}}{2} {e_{-}} ^2$.
This yields a different compatibility  kernel \cite{R13, R15}in  $ \bar{f}_{compat}=\frac{A_1}{2} U(\vec{k}) |e_3(k)|^2$,  as in (A6).

\section{TRANSITIONS WITH $N_V =3$}
We consider two transitions with $N_V=3$ variants, and $N_{OP} = 2$ order parameter  (OP) components, but with different numbers $n$ of non-OP strains. They are the
(a)
cubic/tetragonal transition in 3D  ($n=4$); and (b) triangle/centered
rectangle  in 2D ($n=1$).  See Figs 1 and 2.

\subsection{3D cubic/tetragonal case:\\
 $N_V = 3, N_{OP} = 2, n=4$}

There are three axes
along which the cubic unit cell can elongate, to make a tetragonal cell, so one expects three variants, as in Fig 2.The cubic symmetry group G = $Pm\bar{3}m$ with $\nu_G = 48$ elements goes to the tetragonal group
g = $P4/mmm$
 with $\nu_g =16$ elements, so there are \cite{R6}  $N_V = 3$ variants, as expected.   The variants are generated by joint action of the 
$N_{OP}=2$ order parameters that are the two  3D deviatoric strains\cite{R13,R18,R28},  
with the vector in OP space  chosen as ${\vec e} = (e_3, e_2) \sim (\frac{1}{\sqrt{6}}\{X^2 + Y^2 - 2 Z^2\}, \frac{1}{\sqrt{2}} \{X^2 - Y^2\})$. 
 
The cubic/tetragonal Landau free energy has been considered by Barsch and Krumhansl and others \cite{R18,R28}. The  invariants $\{I_p\}$  under the cubic point group, up to a maximum order $p_{max}=4$, are
$I_2=e_3^2 +e_2^2\equiv\varepsilon^2$, $I_4= I_2^2$, and a  third-order invariant 
$I_3=e_3^3-3 e_2 ^2 e_3$. This is explicitly seen  to be  a scalar under cubic-symmetries,   as it can be written with (2.6a) in terms of  invariants $(XYZ)^2$ and $X^q + Y^q + Z^q$ with $q =2,4,6$. 
From BK scalings  as before,  and sign choices $\sigma_3 = -1, \sigma_4 = +1$ for three minima, we have $\bar{f_L}=(\tau-1){\vec{e}}^2+ f_{0}(e_3,e_2)$, where the temperature-independent $f_0$  of (2.18a) is

$$f_{0}=I_2-C_3I_3+C_4I_4 , (4.1) $$ 

\noindent and the unscaled elastic constants $C^{(3)}$, $C^{(4)}$
are related to the scaled ones as 
$$C_3=C^{(3)}\lambda^3/E_0 ; ~~~C_4=C^{(4)}\lambda^4/E_0 .~~~(4.2)$$

In polar coordinates in OP space \cite{ R6},
with  
$ \vec{e} \equiv (e_3 , e_2)=\varepsilon(\cos\phi,\sin\phi)~$
of magnitude 

$$\varepsilon \equiv |\vec{e}|   = [{e_3}^2 + {e_2}^2]^{1/2}, ~~~~~(4.3)$$
the symmetry in OP space is manifestly carried by the third-order
invariant, 
 
$$I_3=\varepsilon^3(\cos^3\phi-3\cos\phi\sin^2\phi) = \varepsilon^3 \cos 3 \phi.~ (4.4)$$
Then  $f_0$  in the form of (2.23), with $ \eta_3\equiv\cos3\phi$ is

$$f_0 = \varepsilon^2 - C_3 \varepsilon^3  + C_4 \varepsilon^4 + \Delta f_0 ;~~~(4.5a)$$
$$\Delta f_0 =  C_3 \varepsilon^3(1 - \eta_3  ).~~~~ (4.5b)$$

The angular dependence is  in $\Delta f_0 \sim - \cos 3 \phi$. The radial minima and degeneracy  conditions  on $f_{0}$ with $\eta_3 (\phi_m) = 1$, yield $2- 3C_3 + 4 C_4 = 0$ and $1- C_3 + C_4 = 0$,  fixing the two coefficients as

$$C_3=2; ~~~C_4=1,~~~(4.6)$$
achieved  by choosing scaling parameters 

$$\lambda=C^{(3)}/2C^{(4)}; ~ E_0= C^{(4)}(C^{(3)}/ 2C^{(4)})^4. ~(4.7)$$

Then $\bar{f}_L (e_3, e_2)$ and hence $\bar{\varepsilon} (\tau)$ is independent of material constants, i.e. there is quasi-universality of the first kind.
The $N_V=3$ variant minima of $f_0(\varepsilon,\phi)$ are at angles $\eta_3 (\phi_m) = \cos 3 \phi_m =1$, and radius 
$\varepsilon_m = \bar{\varepsilon} (\tau)$, where

$$\sin 3 \phi_m = 0; ~~\phi_m =\frac{2(m-1)\pi}{ N_V}, ~m=1,2,3 (=N_V); ~(4.8)$$
$$\bar{\varepsilon}(\tau)=\frac{3}{4} [1+\sqrt{1-8\tau/9}]. ~(4.9)$$ 
 For $\tau=1$ , $\bar{\varepsilon} (\tau) =1$, as required, and the upper spinodal is  universal, $\tau = \tau_{up} = 9/8$. The saddle-point barriers are at  radius  
$\bar{\varepsilon}_b(\tau)=\frac{3}{4} [1- \sqrt{1-8\tau/9}]$, and angles $\phi_{b m} = (2 m - 1) \pi/3$.

At transition the minima of the Landau free energy fall on the $N_V +1
= 4$ vertices  and center of a rightward-pointing equilateral triangle \cite{R13,R28} inscribed in a unit circle
in $N_{OP} = 2$ dimensional OP space, with corners at $(e_3 , e_2) = (1,0), (- 1/2,  \pm \sqrt{3}/2)$.  (For a different sign choice $\sigma_3 = +1$, the triangle merely changes direction.) See  Fig 3 and Table I. 
The upper bound on the possible types of domain wall between pairs of variants is $N_W = N_V ! /[ 2 ! (N_V - 1)!] =N_V (N_V - 1)/2 =3$.
  
The scaled Landau free energy $\bar{f}_L = (\tau -1) \varepsilon^2 + f_0$  in OP components with the choice (4.6)  is

$$f_0 =  I_2- 2I_3+ I_4 = e_3^2 + e_2^2 - 2(e_3^3-3e_3 e_2^2) + (e_3^2 +e_2^2)^2 $$ 
$$ =  (1+ 2 e_3 ) (3 e_2^2 - e_3^2 +2 e_3 -1) + ({\vec{e}} ^2 - 1)^2 ,~(4.10a) $$ 
where the second equation \cite{R13} explicitly shows the  fourfold degenerate roots  of $f_L =0$ at transition. 

 In polar coordinates as in (2.23), 

$$\bar{f}_L (\varepsilon, \phi) = [ (\tau -1) \varepsilon^2 + \varepsilon^2(\varepsilon-1)^2] + \Delta f_0 ,~~(4.10b)$$
where   $\Delta f_0 \equiv f_0 (\varepsilon, \phi) - f_0 (\varepsilon, \phi_m) = 2(1-\eta_3)\varepsilon^3$ vanishes in the angular directions $\phi = \phi_m$ of the minima. As  $\bar{f}_L (\bar{\varepsilon},  \phi)$ is material-independent,  there is again quasi-universality of the first kind. Here from the definition of (2.22d), $g_L = \tau -1 + (\bar{\varepsilon}-1)^2$.

The Ginzburg term is $\bar{f}_G = \xi^2[( \vec{\Delta} e_3)^2 + (\vec{ \Delta} e_2)^2 ]$.  There are  
$n=4$ non-OP compressional  $e_1$, and shear strains $e_4 , e_5 , e_6$, that can be written in terms of the OP strains, so   $e_i=\sum_{\ell =2,3}B_{i \ell} e_\ell$ with $i = 1, 4, 5, 6$. Substituting into the non-OP harmonic terms yields the cubic/ tetragonal potential 
with the $2 \times 2$ matrix kernel of (2.19), given in (A23) of the Appendix. 
 The 3D  relaxational OP strain simulations  \cite{R14} did not  explicitly
state the kernel, now given here for completeness.

\subsection{Triangle/centred-rectangle  case: $N_V = 3, N_{OP} = 2, n=1$}

There are three ways to convert an equilateral to an isosceles triangle, with the unit cell of the (equilateral) triangle changing to a centered rectangle, so one expects three variants, as in Fig 1.  The OP are 2D $N_{OP}=2$ deviatoric and shear strains, with the OP vector chosen as  ${\vec e} = (e_2, e_3) \sim  \frac{1}{2} (X^2 - Y^2, XY)$.
The triangle point group G= $p6mm$ with $\nu_G = 6$ elements
goes to the centred rectangle subgroup g = $c2mm$
 with $\nu_g =2$ elements, so there are \cite{R15} $N_V = 3$ variants, as expected. The third-order invariant under $\pi/3$ rotations of the triangular lattice
 is now  $I_3 = e_2 ^3 - 3 e_2 e_3 ^2$. For $(X, Y) = R( \cos \alpha, \sin \alpha)$, one finds $I_3 = (R^3 /8) \cos 6 \alpha$,  manifestly invariant under the 
 $\alpha \rightarrow \alpha + \pi /3$ triangular symmetry. The scaling parameter choices are as before.
 The final scaled Landau free energy  \cite{R13} is formally similar to  the cubic/tetragonal case of (4.10a), with $e_2$ and $e_3$ interchanged,
$$f_0 =  I_2- 2I_3+ I_4 = e_2^2 + e_3^2 - 2(e_2^3-3
e_2 e_3^2) + (e_2^2 +e_3^2)^2. ~(4.11)$$ 

At transition the $N_V + 1 = 4$ degenerate Landau minima again fall
 on the three vertices and at the center of an equilateral triangle
 inscribed in a unit circle. The maximum number of domain wall types is $N_W = 3$.

The Ginzburg terms are the same as above, while the compatibility potential in 2D, is of course different.  The  single $n = 1$ non-OP (compressional)  strain is $e_1 =- \sum_{\ell = 2,3} O_\ell e_\ell / O_1$ from the 2D compatibility constraint. Substitution  into the harmonic term as in (2.19) immediately yields \cite{R13,R15}, $\bar{f}_{non}= (A_1/ 2) |e_1 (\vec k)|^2 \rightarrow \bar{f}_{compat} (e_2 , e_3)$ with a
  $2 \times 2$ compatibility-kernel matrix of (A2).

\section{TRANSITIONS WITH $N_V=4$}

We consider two transitions with $N_V=4$ variants, and different order parameter (OP) components. They are the
(a) cubic/trigonal case in $3D$ ( $N_{OP}=3$, $n=3$); and (b)
square/oblique case in $2D$ ($N_{OP}=2$, $n=1$). See Figs 1 and 2.

\subsection{Cubic/trigonal case: $N_V = 4, N_{OP} = 3, n=3$}

The distortion acts  along body diagonals of the cube and  the $N_{OP}=3$ shears  are the three components of the OP vector $\vec{e} = (e_4, e_5, e_6) \sim (YZ, ZX, XY)$.  With the cubic group G = $Pm\bar{3}m$ with $\nu_G =48$ going to the trigonal or rhombohedral subgroup g = $P\bar{3}1m$ with $\nu_g = 12$,
there are \cite{R6} $N_V=4$ shear-induced variants. 
The four  invariants up to order $p_{max} = 4$ are   $I_2=
e_4^2+ e_5^2+ e_6^2; ~I_3= e_4 e_5 e_6;~
I_4=e_4^4+ e_5^4+ e_6^4;$ and $I'_4 = {I_2}^2$, with $N_{mat} = 4$ material coefficients. The scaled Landau free energy  is

$$\bar{f_L}=(\tau-1)I_2 +f_{0}(e_4, e_5, e_6) , ~~~(5.1a)$$

where

$$f_{0}=I_2 - C_3 I_3+ C'_4 I_2^2+C_4I_4 ,~~~~(5.1b)$$
and the scaled parameters are related to unscaled ones by

$$C_3 = \lambda^3 \frac{C^{(3)}}{ E_0} ~; C'_4 = \lambda^4 \frac{C'^{(4)}}{E_0} ~; C_4 = \lambda^4 \frac{C^{(4)}}{E_0} .~~(5.2)$$

In spherical polar coordinates in OP space, 
$$\vec{e} = (e_4, e_5, e_6) =\varepsilon(\sin\theta \cos\phi,\sin\theta\sin\phi,\cos\theta),~~(5.3a)$$
with magnitude
$$\varepsilon \equiv |\vec{e}| = [ {e_4}^2 + {e_5}^2 + {e_6}^2]^{1/2}. ~~~(5.3b)$$
The invariants  in OP space are then

$$I_2=\varepsilon^2; ~~~I_3=\varepsilon^3  \sin^2 \theta \cos \theta \sin \phi \cos \phi; I' _4  = \varepsilon^4 ,$$
$$I_4=\varepsilon^4[\sin^4\theta(\cos^4\phi+
\sin^4\phi)+\cos^4\theta]. ~(5.4)$$

Using trigonometric identities,  $f_0$ is
 
$$f_0 (\varepsilon, \theta, \phi ) = \varepsilon^2 +C'_4 \varepsilon^4  -\frac{ C_3}{4} \varepsilon^3 \sin\theta \sin 2\theta \sin 2\phi  $$
$$+ C_4 \varepsilon^4 [ 1 - \frac{1}{2} \sin^2 2 \theta - \frac{1}{2} \sin^4 \theta \sin^2 2 \phi ] , ~(5.5)$$
and has three remaining material constants $C_3, C_4, C'_4$, while there are two scaling parameters $\lambda, E_0$.

The radial minimum $\partial f_{0}/\partial\varepsilon +2 (\tau - 1) \varepsilon = 0$, and $\tau = 1$ degeneracy condition $f_{0}(\varepsilon=1,\theta_m, \phi_m)=0$ yield

$$C_3 = 6 \sqrt{3};~ {C_4}' = 1 - C_4  /3,~~~~~ (5.6)$$
achieved by the choice of scaling parameters 

$$\lambda =\frac{(C^{(3)}/6 \sqrt{3})}{[{C'}^{(4)} + (C^{(4)} /3) ]} ;~~~
E_0=\ \lambda^3 \frac{C^{(3)}}{6 \sqrt{3}}.~~ (5.7).$$

The angular minimum conditions $\partial f_0/\partial \phi =0$, and $\partial f_0 /\partial \theta = 0$ yield minima at $\cos 2 \phi_m =0$ or at

$$\sin 4 \phi_m = 0,~~~~ 3 \cos^2 \theta_m = 1,~~~~(5.8a)$$

$$\phi_m = (2 m - 1) \pi / N_V , ~ m = 1, 2, 3, 4 (= N_V).~~(5.8b)$$
For appropriate positive signs of the second derivatives, there are two points each on the northern and southern unit hemispheres in OP space,

$$\phi_m = \pi/4 , 5 \pi/4 ,~\theta_m = \bar{\theta};~~$$
$$\phi_m = 3 \pi/4 , 7 \pi/4 ,~\theta_m = \bar{\theta} + \pi, ~~~~(5.9)$$
where $\cos \bar{\theta} = 1/\sqrt{3}$.
These four  minima $m = 1,2,3,4$ are  at the ends of vectors in three-dimensional OP space $\vec{e} = \vec{e}_m \equiv \varepsilon_m (\sin \theta_m \cos \phi_m, \sin \theta_m \sin \phi_m, \cos \theta_m)$, that
have equal relative separations or polyhedral sides  of e.g. $ |\vec{e_1} -\vec{e_2}| =  2 \sqrt{2/3}~ \varepsilon_m$, and relative cosines  of eg $\vec{e_1}.\vec{e_2}/\varepsilon_m ^2 = \cos \psi = -1/3$, so the angle between vectors
is the well-known  tetrahedral angle $\psi = \cos^{-1} (-1/3) = 109^{o} 28'$.

At transition, the Landau minima fall on the vertices and center of a tetrahedron,
 inscribed in a sphere of unit radius in $N_{OP} = 3$ dimensional order parameter space, as in  the schematic points of  Fig 3. The generic numbers for transition are in Table I.   The maximum number of domain wall types is $N_W =6$. Of course in this and other cases, this is only an upper bound, and energetic considerations might lead to fewer types actually appearing in the final microstructure.

Then the scaled Landau free energy  in components  is

$$\bar{f_L} = (\tau -1) I_2 +\{ I_2 - 6 \sqrt{3} I_3 + (1 -  C_4 /3) I_2^2 + C_4 I_4\} .~(5.10a)$$

\noindent In polar coordinates as in (2.23), it  is

$$\bar{f}_L=[(\tau-1)\varepsilon^2+ \varepsilon^2 (\varepsilon - 1)^2] + \Delta f_0 , ~~~(5.10b)$$

\noindent manifestly showing the degeneracy  at transition $\bar{f}_L (1, \theta_m, \phi_m) =0$. The angular part 
$\Delta f_0 \equiv f_0 (\varepsilon, \phi, \theta) - f_0(\varepsilon, \phi_m , \theta_m)$, is

$$\Delta f_0 = \frac{3 \sqrt{3}}{4} \varepsilon^3 [ \frac{4}{3 \sqrt{3}}  -\sin \theta \sin 2 \theta  \sin2\phi ]$$
$$+ C_4 \varepsilon^4 [ \frac{2}{3} - \frac{1}{2} \sin^2 2 \theta - \frac{1}{2} \sin^4 \theta \sin^2 2 \phi ]. ~~~(5.11)$$
This carries  material dependence through $C_4$, but vanishes at minima, so $\bar{f}_L (\varepsilon, \theta_m , \phi_m)$  and hence
$\bar{\varepsilon} (\tau)$  are still universal for all $\tau$ i.e. there is quasi-universality of the second kind. Here from (2.22d),  $g_L = \tau -1 + (\bar{\varepsilon} - 1)^2$.

The material-independent OP magnitude at the tetrahedron corners  is 

$$\bar{\varepsilon}(\tau)=\frac{3}{4} [1+\sqrt{1-8\tau/9}],~~~(5.12)$$ 
and happens to be the same form as for the cubic/tetragonal transition, with the saddle-point barrier at $\bar{\varepsilon_b}(\tau)=\frac{3}{4} [1- \sqrt{1-8\tau/9}]$.
The Ginzburg term is $\bar{f}_G = \xi^2 [\sum_{\ell =  4, 5, 6} ({\vec \Delta} e_\ell)^2 ]$. The $n=3$ non-OP strains are  the compressional
 ($e_1$) and the deviatoric strains ($e_2, e_3$). The $3 \times 3$ compatibility kernel 
$U_{\ell \ell'}(\vec k)$ of (2.19) is given in (A20).

\subsection{Square/oblique case : $N_V = 4, N_{OP} = 2, n=1$}

A square converts to an unequal-sided oblique polygon  with the deviatoric $e_2$, and shear
$e_3$ order parameters acting simultaneously. Since each can distort the square in two ways, one expects four variants, as in Fig 1. The harmonic parts are
assumed to soften at the same $T_c$ (otherwise they would be two
separate, unrelated transitions). The joint action of the two order parameters is through their coupling.
The square point group G= $p4mm$ with $\nu_G = 4 $ goes to
g = $p2$ with $\nu_g = 1$, so there are \cite{R15}  $N_V =4$ variants, as expected. The anharmonic invariants
are as in the square/rectangle case, $e_2^4$, $e_2^6$ and $e_3^4$, $e_3^6$ or $p_{max} =6$ separately for each OP. We consider the simplest
case of equal elastic constants, and the simplest coupling ${e_2}^2e_3^2$.
 The scaled free energy  term $f_0$ is:

$$ f_0 = (e_2^2+ e_3^2)-C_{4}(e_2^4 + e_3^4)+ C_{6}(e_2^6 + e_3^6)] - C' _4 {e_2}^2 {e_3}^2 ,~(5.13)$$
where the scaled and unscaled  coefficients are related  by 
$C_4=C^{(4)}\lambda^4 /{E_0}; ~~~C_6= C^{(6)}\lambda^6/{E_0};
~~~C'_4= {C'}^{(4)}\lambda^4 / {E_0}.$

Transforming to polar coordinates $(e_2, e_3)=\varepsilon(\cos\phi , \sin\phi)$, with  $\eta_2\equiv\cos2\phi$,  and using trigonometric identities, this becomes

$$f_{0}=  t_0 (\varepsilon) + t_4 (\varepsilon)  {\eta_2}^2 ,~~~(5.14)$$ 
where the coefficients  are
$$t_0\equiv \varepsilon^2 -\frac{\varepsilon^4}{4}(2 C_4 +C'_4 )
+\frac{\varepsilon^6}{4} C_6 ,~~~ (5.15a) $$ 

$$t_4\equiv \frac{3}{4} C_6 \varepsilon^6 -\frac{(2C_4-C'_4) \varepsilon^4}{4}.~~~~~(5.15b)$$

The angular dependence is $f_0 \sim   \cos 4 \phi$. The degeneracy condition, and the radial minimum condition at transition finally yield $C_4+C'_4 /2 =4;~~C_6=4$, achieved through the choice

$$\lambda^2= ( C^{(4)} +  {C'}^{(4)} /2) / C^{(6)};~E_0= \lambda^6 C^{(6)} /4. ~(5.16)$$

The angular minima are at  ${\eta_2 (\phi_m)}^2 = 1$ or

$$\sin4\phi_m=0; ~\phi_m=\frac{(2m-1)\pi}{N_V},~ m= 1,2,3, 4 (=N_V).~(5.17)$$

Then  the scaled free energy in OP components is 

$$ f_0 = {\vec e}^2- (4 - \frac{1}{2}C'_4 )(e_2^4 + e_3^4)+ 4(e_2^6 + e_3^6) - C' _4 {e_2}^2 {e_3}^2.~(5.18a)$$

In polar coordinates as in (2.23), 

$$\bar{f} _{L}=[(\tau-1)\varepsilon^2+\varepsilon^2(\varepsilon^2-1)^2] + \Delta f_0, $$
$$\Delta f_0 = \varepsilon^4 (3\varepsilon^2- 2 + C'_4 /2 )\cos^2 2\phi , ~~(5.18b)$$
so the quasi-universality is of the second kind, with $\bar{f}_L (\varepsilon, \phi_m)$ and
 $\bar{\varepsilon} (\tau)$ of (3.6), both independent of material constants. Here $g_L = \tau - 1 + ({\bar \varepsilon}^2 - 1)^2$.

 At transition, the $N_V =4$ variant  minima fall on the
 vertices of a square inscribed in a unit circle in $N_{OP}=2$
 dimensional order parameter space.  These are the only variants  for $C'_4 >1$, as the trivial roots on the axes $(e_2 , e_3) = (\pm {\bar \varepsilon}, 0), (0, \pm {\bar \varepsilon})$  are then unstable. The maximum possible number of domain wall types is $N_W = N_V (N_V -1)/2 =6$.

The Ginzburg term is $f_{G}=\xi^2[(\vec{\Delta}
e_2)^2+(\vec{\Delta} e_3)^2]$. There is a single  $n=1$ non-OP variable $e_1$ as in the triangle-centered rectangle case, so   
the compatibility  kernel from the compressional harmonic term $e_1 ^2$  is the same  as the $2 \times 2$ matrix $U_{\ell \ell'}$ of (A2).

\section{Transitions with $N_V =6$}

We consider two transitions with $N_V=6$ variants, and $N_{OP} = 2$ order parameters (OP), but with different numbers $n$ of non-OP strains. They are the
(a)
cubic/orthorhombic case in 3D  ($n=4$); and (b) triangle/center
rectangle case in 2D ($n=1$).  See Figs 1 and 2.

\subsection{Cubic/orthorhombic case $N_V =6, N_{OP} = 2, n=4$}

For a cubic to orthorhombic distortion, the cross-sectional area perpendicular to each of  three axes
can be rectangular in two ways, so one expects six variants, as in Fig 2. The symmetry group G = $Pm\bar{3}m$ with $\nu_G = 48$ elements goes to the orthorhombic group
g = $Pmmm$
 with $\nu_g =8$ elements, so \cite{R6} $N_V = 48/8 =6$, as expected. The variants are generated by combined action of the 
$N_{OP}=2$ order parameter components that are the two  3D deviatoric strains,  
with the vector in OP space  chosen as in the cubic/ tetragonal case, ${\vec e} = (e_3, e_2) \sim (\frac{1}{\sqrt 6}\{ X^2 + Y^2 - 2 Z^2\}, \frac{1}{\sqrt 2} \{X^2 - Y^2 \})$. 
 
The cubic/orthorhombic free energy in Cartesian strains has been considered for fitting to  FePd experiments \cite{R7}; here however, we work with physical strains. The previous cubic/tetragonal case of (4.1),  with a third order invariant $I_3 ={ e_3}^3 - 3 e_3{ e_2 }^2$,  yielded three minima alternating with three maxima, on the unit circle. For six minima on the unit circle,  a sixth order invariant $I_6 = I_3 ^2$ will be the leading angular term. 

We consider two cases, with up to sixth order, and  up to eighth order invariants. For  invariants of up to $p_{max} = 6$th
order, the free energy has  $N_{max} =4$ material coefficients. 
$\bar{f}_L=(\tau-1)( e_3^2+e_2^2) +f_{0}
(e_3, e_2)$ where $f_{0}$ is

$$f_{0}=I_2-C_4I_2^2-C_6I_6+ C'_6I_2 ^3 ,~(6.1a)$$
with signs $\sigma_4 = \sigma_6 = -1, \sigma' _6 = +1$.
 For materials with other coefficient signs,  we are forced to go to higher $p_{max} = 8$th order, and  
the additional invariants  are  $I_8= I_3^2I_2 , I_8 ' = I_2 ^4$. (The odd invariants  $I_3, I_5= I_3I_2,
I_7=I_5I_2$ give sign-varying contributions to derivatives  $\partial f_0 /\partial \phi$ at different minima that should be equivalent,  so we set their coefficients to zero from the start.)

The scaled free energy up to eighth order, with  $N_{max} =6$ material coefficients is 

$$f_{0}=I_2-C_4I_2^2+C_6I_6-C'_6I_2^3-C_8I_8+C'_8I_2^4 .~(6.1b)$$ 
It is convenient to define  ${C_6}^{(-)} \equiv C'_6 - C_6, {C_8}^{(-)} \equiv C'_8 - C_8$.
Transforming to polar coordinates, ${\vec e} =(e_3,e_2)=
\varepsilon(\cos\phi,\sin\phi)$  we get  $I_3= \varepsilon^2\cos3\phi$ as before, so
$I_6=\varepsilon^6\eta_3^2, I_8=\varepsilon^8 \eta_3^2$  where
$\eta_3\equiv\cos3\phi$. 

Collecting terms, the sixth and eighth order cases of  (6.1a) and (6.1b) can both be written as 
$$f_{0}=t_0 (\varepsilon) + t_6 (\varepsilon) ( 1- \eta_3^2),~~(6.2)$$
where from (6.1a)
$$t_0\equiv \varepsilon^2 -C_4\varepsilon^4+{C_6}^{(-)}\varepsilon^6;
~t_6\equiv  C_6 \varepsilon^6;~(6.3a)$$
while from (6.1b),
$$t_0\equiv \varepsilon^2 -C_4\varepsilon^4-{C_6}^{(-)}\varepsilon^6+ {C_8}^{(-)} \varepsilon^8;
~t_6\equiv C_8\varepsilon^8 - C_6 \varepsilon^6.~(6.3b)$$
In both cases, the angular dependence is $f_0 \sim - \cos 6 \phi$.

For the sixth order case of (6.3a), the
degeneracy $f_{0}(\varepsilon=1,\phi= {\phi}_m)=0$ and
radial minimum condition $\partial f_{0}/\partial\varepsilon +2 (\tau - 1) \varepsilon=0$
determine two of the constants as
${C_6}^{(-)}= 1 , C_4 = 2$. The scaling parameters for these values are, with unscaled elastic-coefficient ratio $\alpha \equiv C^{(4)}/ ({C' }^{(6)}  - C ^{(6)})$,

$$\lambda^2 =\alpha /2 , ~~~~E_0=\lambda^4 C^{(4)} /2 .~(6.4)$$

Then the angular contributions clearly yield  $\partial f_0 /\partial \phi = 0$ roots at 

$$\sin6 \phi_m=0;~{\phi}_m=\frac{2(m-1)\pi}{N_V}, ~
m=1,...6 (=N_V).~(6.5)$$
where ${\eta_3 (\phi_m)}^2 = 1$.

The scaled free energy in  components in OP
space is, with the above scaled coefficients,

$$\bar{f}_L =(\tau-1)I_2  + I_2 (I_2 - 1)^2  +  C_6( {I_2}^3 - {I_3}^2 ). ~~(6.6)$$

In polar coordinates as in (2.23),

$$\bar{f_L} =[ (\tau-1)\varepsilon^2+\varepsilon^2(\varepsilon^2 - 1)^2  ] +\Delta f_0,$$

$$\Delta f_0 =  \frac{1}{2} C_6 \varepsilon^6 ( 1 - \cos 6 \phi ) ,~~~~(6.7)$$ 
The last term  $\Delta f_0$ vanishes at the six minimal directions,  where the material constant $C_6$ is eliminated,  so there is quasi-universality of the second kind. The  OP magnitude  ${\bar \varepsilon}(\tau)$ is as in the tetragonal/orthorhombic case of (3.6).                                                             
The variants are manifestly degenerate with austenite as $\bar{f_L} (1, \phi_m) =0$ at transition. 

For the eighth order case of (6.3b), there are  three material constants $C_4, {C_6}^{(-)}, {C_8}^{(-)} $ in $\bar{f_L} (\varepsilon, \phi_m)$,  and only two remaining scaling parameters                    $\lambda, E_0$. 
The degeneracy and
radial minimum condition now 
determine two of the constants as
${C_6}^{(-)}= 3 - 2 C_4 ,{C_8}^{(-)}= 2 - C_4$. The equivalent condition $3 {C_8}^{(-)} - 2 {C_6}^{(-)} -C_4 = 0$,  yields with (2.12), a quadratic, $ 3\lambda^4 - 2\gamma \lambda^2 - \alpha \gamma$, where $\alpha$ is as above and $\gamma \equiv  ({C' }^{(6)}  - C ^{(6)})/ ({C' }^{(8)}  - C ^{(8)})$. The positivity of second derivatives requires that $C_8 > C_6$
and $2 - C_4 > 0$ while $\lambda^2 >0$ below,  further requires $3 - 2 C_4 > 0$.

However $\lambda  (\ll 1)$ can also be obtained from the relation between scaled and unscaled coefficients (2.12),  as $\lambda^2 = \alpha (3 - 2 C_4)/C_4 = \gamma (2 -C_4)/ (3- 2C_4)$.
Demanding consistency yields $\lambda, E_0, C_4$  in terms of the unscaled  elastic coefficients, but here $C_4$ is no longer just a universal number. The scaling parameters are then

$$\lambda = (\gamma /3) [ 1 +\{1 + 3  \alpha/ \gamma\}^{1/2} ] ; ~~ E_0 = C^{(4)} \lambda^4  / C_4 ;~~(6.8a)$$
$$C_4 = (3/2) / [ 1 +(\gamma / 6 \alpha)(1 + \{1 + 3  \alpha/ \gamma\}^{1/2}) ] .~~ (6.8b)$$

As elastic constants vary, the constant $C_4 (\alpha/\gamma)$ moves in a narrow range $3/2 > C_4 > 0$, e.g.  $C_4 = 1$ for $\alpha /\gamma = 1$.

The scaled free energy in  OP components  for the eighth order case  from (6.3b) is then

$$\bar{f}_L =(\tau-1)I_2  + I_2 (I_2 - 1)^2 \{ 1 + (2 -C_4) I_2 \} $$
$$ + (C_8 I_2 - C_6)( {I_2}^3 - {I_3}^2 ). ~ ~~(6.9)$$
In polar coordinates as in (2.23),

$$\bar{f_L} =[ (\tau-1)\varepsilon^2+\varepsilon^2(\varepsilon^2 - 1)^2 \{ 1 + (2- C_4) \varepsilon^2\} ] +\Delta f_0,$$

$$\Delta f_0 = \frac{1}{2} (C_8\varepsilon^8 - C_6 \varepsilon^6 )( 1 - \cos 6 \phi ) .~~~~(6.10)$$ 
with $C_8 , C_6$ eliminated  at minimal angular directions.
                                                               
The  root $\bar{\varepsilon} (\tau)$ is
 the solution of a cubic
$$4(2- C_4)X^3+ 3 ( 5 - 2 C_4)  X^2+  2 (3 - C_4)X+(\tau-1)=0, ~~(6.11)$$
where $X\equiv\overline{\varepsilon}^2 -1 \geq 0$. At $\tau = 1$, the required root is $X =0$, 
and just below transition is linear, $X \simeq (1 - \tau)/2 (3 - C_4)$. Close to zero temperature, with $\tau = -|\tau|$
and for $|\tau(T=0)| = T_c /(T_0  - T_c) >> 1$, one has $X \simeq [ |\tau|/4 (2 - C_4) ]^{1/3}$.
Thus the scaling procedure carries through even for the eighth order case, but there is now quasi-universality of the third kind, with a weak residual  material-dependence through $C_4$.

  At transition and for both cases, the $N_V = 6$ nontrivial Landau minima fall on the
  vertices of a hexagon \cite{R7} inscribed in a unit sphere in $N_{OP}=2$
  order-parameter space as in Fig. 3 and Table I. The upper bound on the number of possible martensite-martensite domain wall types is $N_W = N_V (N_V - 1)/2 =15$, although as mentioned, not all of these may be seen, for energetic reasons.

The Ginzburg term, and the compatibility potential from  the $n=4$ non-OP harmonic term is the same
as in the cubic/tetragonal  case of (A23) of the Appendix.

\subsection{Triangle/oblique case $N_V =6, N_{OP} = 2, n=1$}

The two order parameter components are the single deviatoric and single shear strains,    ${\vec e} =(e_2, e_3) = \varepsilon (\cos \phi,\sin \phi)$.
The triangle point group G = $p6mm$ with $\nu_G = 6$ goes to the subgroup g =$p2$  with $\nu_g = 2$ so there are \cite{R15}
$N_V = \nu_G/\nu_g=6$ variants. (The ratio of the numbers of rotational elements  would however give $3/1 =3$ variants.)

The scaled Landau free energy density \cite{R13}
 is formally similar to  the cubic/orthorhombic case of (6.7). 
At transition the $N_V + 1 = 6 +1$ degenerate Landau minima again fall
 on the six vertices and at the center of a hexagon
 inscribed in a unit circle, as in  Fig 2. The maximum number of domain wall types is $N_W = 15$, but not all may finally appear.

The Ginzburg term, and the compatibility potential from  the $n=1$ non-OP harmonic term is the same
as in the triangle/centered rectangle case of (A2).

 \section{PSEUDOSPIN HAMILTONIANS}
 
        The idea of using discrete-variable pseudospins to approximate continuous-variable  distortions on a lattice  was proposed earlier \cite{R20,R21}, and has been pursued \cite{R22,R23,R24,R25}. We had suggested obtaining 
  pseudospin hamiltonians for various transitions by substituting the order parameter (OP) values at the polar coordinate Landau minima, into the total scaled free energy \cite{R22}.
 The $N_{OP}=1$ case for the square/rectangle case as $e_2 (\vec r) \rightarrow  \bar{\varepsilon} S (\vec r)$, where the three minima induce a spin-1 model $S= 0, +1, -1$ with  three values on a line, of a single-component   pseudospin. 
 With $S^6 = S^4 = S^2 = 1$ or $0$, the nonlinearities in the Landau free energy collapse to 
 $\bar{f}_L  \rightarrow \bar{\varepsilon} (\tau)^2 g_L (\tau) S^2$ where $g_L (\tau) = \tau - 1 + (\bar{\varepsilon}^2 - 1)^2$ changes sign at transition. The Ginzburg and compatibility terms are also written in terms of pseudospins. This yields a temperature-dependent pseudospin hamiltonian $H(S) \equiv F ( e_2 \rightarrow {\bar \varepsilon} S)$ that is like a generalized Blume-Capel spin-1 model \cite{R19}.  The hamiltonian has a temperature-dependent quadratic term, a nearest-neighbor ferromagnetic  term, and a PLA compatibility term \cite{R23},
 
$$ \beta H(S) = \frac{D_0}{2}[ \sum_{\vec r}  \{g_L S^2 (\vec r) + \xi^2 ( \vec{\Delta} S)^2\}$$
$$ +  \sum_{{\vec r}, {\vec r'}} \frac{A_1}{2} U({\vec r} - {\vec r'}) S(\vec r) S(\vec r') ],~~~(7.1) $$
with gradient $\vec{\nabla}$  realized as difference operators  $\vec{\Delta}$ on a computational grid, as mentioned.
Here  $D_0 (T) \equiv 2 \bar{\varepsilon}^2 (\tau) E_0 /k_B T$, and the hamiltonian is diagonal in Fourier space,

$$\beta H = \frac{1}{2} \sum_{\vec k}   Q_0 (\vec k)|S (\vec k)|^2 ;~~~~(7.2a)$$

$$Q_0  (\vec k) \equiv D_0 [ g_L (\tau) +  \xi^2 {\vec K}^2   + \frac{A_1}{2} U (\vec k )] .~~(7.2b)$$
Here on a grid of unit lattice constant, $K_\mu\equiv 2 \sin (k_\mu /2)$, with $\mu = x, y$.
This hamiltonian has been studied in a local meanfield approximation, under a cooling ramp obtaining glassy domain-wall textures, dependent on cooling rate and initial conditions \cite{R23}.

 For $N_{OP} >1, N_V > 2$ transitions we {\it do not} simply  get a generalized spin-$j$ model with $2j +1$ states on
 a line, where $j = N_V/2$. Instead we obtain clock-like models  \cite{R19} with discrete $\vec{S}$ vector variables
 pointing to $N_V +1$ corners  and centre of a polyhedron in  $N_{OP}$-dimensional space, as denoted by arrows in Fig. 3. Since the zero state is included, these  may be termed `clock-zero'  $\mathbb{Z}_{N_V +1}$ models. Note that, unlike pure clock $\mathbb{Z}_N$ models, the spin square-magnitude ${\vec{S}}^2(\vec r)$ is still a statistical variable and not a constant, because of the zero states.  
 Choosing $N_{OP} = 2,3$ component strains only at the minima induces vector pseudospins in OP space,
  
 $${\vec e}  (\vec r)\rightarrow \bar{\varepsilon} {\vec S} (\vec r) .~~ (7.3)$$
 The variant angular dependence ${\bar f}_L \sim - \cos N_V \phi$ generates the clock-variable directions. 
 
 The general  temperature-dependent pseudospin hamiltonian is 
 
$$H(S_\ell (\vec r)) = { F}(e_\ell \rightarrow  \bar{\varepsilon} S_\ell).~~(7.4)$$
As in the square/rectangle case, the radial part of the Landau term with OP nonlinearities, collapses to a quadratic in the pseudospin magnitude, since $n$th  powers  of the spin-vector magnitude $|\vec{S}|^n = |\vec{S}|^2 = 0, 1$.

Although in zero stress  the uniform state is no longer a Landau minimum  below the lower spinodal $T_c$, there is a possibility that nonuniform textures  exert local internal stresses to favor the zero value at a site even at  low temperatures.  Also, the  original free energy in OP strain always has a turning point at the origin to support dynamical transient zeros, that although few in number, could play a catalytic role in microstructural evolution \cite{R23}. Hence we retain zero spin values at all temperatures, allowing their permanent/transient existence to be determined dynamically. 

The hamiltonian  in coordinate space is

$$\beta H = \frac{D_0}{2}[  \sum_{{\vec r} , \ell} \{ g_L(\tau) S_\ell (\vec r)^2 +    \xi^2 (\vec{\Delta} S_\ell) ^2   \}$$
$$+   \sum_{{\vec r}, {\vec r'}, \ell, \ell'} \frac{A_1}{2} U_{\ell  \ell'} ({\vec r} - {\vec r'} ) S_\ell (\vec r) S_{\ell'} (\vec r') ],~(7.5a)$$
and is transition-specific  through the  $\vec S$ values, the temperature dependence of $g_L (\tau)$, and the compatibility potential. 
Note that the anisotropic terms $\vec{S}(\vec r).{\overleftrightarrow U}(\vec{r} - \vec{r'}).\vec{S}(\vec r')$ are from compatibility  anisotropies in OP space, and differ from models with electric dipoles $\vec{d}$ that have anisotropies  relative to coordinate space axes\cite{R24} $\{\vec{d}(\vec r).\hat{r}\}\{\vec{d}(\vec r').\hat{r'}\} /|\vec{r} - \vec{r'}|^d$,  (although the powerlaw fall-offs with exponent $d$, are the same).

The hamiltonian is diagonal in Fourier space,

$$\beta H = \frac{1}{2} \sum_{\vec k} \sum_{ \ell, \ell'}  Q_{0, \ell \ell'} (\vec k) S_\ell (\vec k) S_{\ell'} (\vec k)^* ;~~(7.6a)$$

$$Q_{0, \ell \ell'} (\vec k) \equiv D_0 [\{ g_L(\tau) +  \xi^2 { \vec K}^2  \} \delta_{\ell, \ell'} + \frac{A_1}{2} U_{\ell  \ell'} ({\vec k} )].~(7.6b)$$
Transitions with free energy quasi-universality of the first and second kind, have reduced pseudospin hamiltonians with {\it universal}  coefficients $g_L$ of the on-site term. Apart from the overall $E_0$, the  material-dependence is only through the texture-inducing Ginzburg and St Venant terms.
 
For two-component OP cases  in addition to $\vec S =0$,  the pseudospin is
${\vec S} = (\cos \phi_m , \sin \phi_m)$, where $m$ takes on $N_V$ values. Thus for the three variants of the
 cubic/tetragonal and triangle/center-rectangle transition, there are three values $\phi_m = 0, 2 \pi/3, 4 \pi/3$ on
 corners of a triangle, and   $g_L = \tau -1 +  ( {\bar \varepsilon} - 1)^2$ where $\bar{\varepsilon} (\tau)$ is
 from (4.9). The four spin vectors are 
$\vec S = (0,0), (1,0), (-\frac{1}{2}, \pm \frac{{\sqrt 3}}{2})$. 

For the square/oblique transition there are four values $\phi_m = \pi/4, 3\pi/4, 5 \pi/4, 7 \pi/4$ on corners of a square, and  $g_L = \tau -1 + ({\bar \varepsilon}^2- 1)^2 $, with ${\bar \varepsilon}(\tau)$ of (3.6). The five spin vectors  are $(0,0), (\pm \frac{1}{{\sqrt 2}}, \frac{1}{{\sqrt 2}}), (\pm \frac{1}{{\sqrt 2}}, - \frac{1}{{\sqrt 2}})$. 

For the three-component OP of the cubic/trigonal transition, the pseudospin vectors are {\it non-planar} as
${\vec S} = (\sin \theta_m \cos \phi_m, \sin \theta_m  \sin \phi_m, \cos \theta_m)$,  where $\theta_m, \phi_m$ take on four values of (5.8) at the corners of a tetrahedron. The coefficient $g_L = \tau - 1  +  ( {\bar \varepsilon} - 1)^2$, with $\bar{\varepsilon} (\tau)$  the same as the cubic/tetragonal case of (4.9). The five spin vectors are 
$(0, 0, 0), (\pm \frac{1}{{\sqrt 3}}, \pm \frac{1}{{\sqrt 3}},  \frac{1}{{\sqrt 3}}), (\pm \frac{1}{{\sqrt 3}}, \mp \frac{1}{{\sqrt 3}}, - \frac{1}{{\sqrt 3}})$. 

For the two-component OP and six variants of the cubic/orthorhombic and triangle/oblique transitions, $\phi_m = 0, \pi/6, 2\pi/6,...,5\pi/6$ on the six corners of a hexagon, and the seven spin vectors are $(0, 0), (\pm 1, 0), (\frac{1}{2}, \pm \frac{{\sqrt 3}}{2}), (-\frac{1}{2}, \pm\frac{ {\sqrt 3}}{2})$, where   for the $p_{max} =6$ case,  $g_L (\tau) = \tau - 1 + (\bar{\varepsilon}^2 - 1)^2$ is universal, and       $\bar{\varepsilon} (\tau)$  as in (3.6). For the $p_{max} = 8$ case, with quasi-universality of the third kind,  there is  a weak material-dependence  through                                                           $g_L = \tau -1 + ({\bar \varepsilon}^2 - 1)^2 [ 1 + {\bar \varepsilon}^2 (2 - C_4) ]$  with $3/2 > C_4 > 0$, and  $\bar{\varepsilon} (\tau)$ as in (6.11). 

The pseudospin reduced hamiltonians here are not just written down, but are induced from the scaled free energy, that encodes the specific symmetry, nonlinearity, and compatibility of each ferroelastic transition: a continuum-variable materials science model is mapped to a discrete-variable  statistical mechanics hamiltonian.

\section{SIMULATIONS OF STRAIN TEXTURES}
Spatially varying strain textures in ferroelastics can be
numerically simulated  in continuous strains by free energy
relaxations, or by  discrete-strain pseudospin hamiltonians.

\subsection{Scaled free energy relaxations}

Of course, there has been much simulation work in the displacement representation or with phase fields \cite{R28}, and one common problem is the choice of the many $N_{mat}$ material coefficients, that require fitting to experiment for each material \cite{R7}.
With scaled free energy strain dynamics or with pseudospin hamiltonians, the material-dependence is essentially eliminated or reduced, to group  many materials with the same structural transition in the same quasi-universality class. Thus  one does not have to explore the full many-parameter space of  $N_{mat}$ unscaled coefficients, and  simulations by different groups can be  more easily compared.

Local equilibrium microstructures can be found from relaxational
dynamics of unscaled order parameter (OP) strains, $\{e'_\ell(r, t)\}$  and the unscaled free energies:
$\partial e'_\ell /{\partial t}=-\Gamma^{(0)} \partial F /{\partial e'_\ell} ,$
where $\Gamma^{(0)}$ is a kinetic constant of dimension inverse-energy $\times$ time. Scaling strains,  $e'_\ell =  \lambda e_\ell$ and energies 
$F = E_0 \bar{F}$, where $\bar{F} =\bar{F}_L +\bar{F}_G +\bar{F}_{compat}$, we obtain a characteristic decay rate $\Gamma_0 \equiv \Gamma^{(0)} E_0 /\lambda^2$. Then with a dimensionless time $\bar{t} \equiv \Gamma_0  t$ absorbing  $E_0$, 
a scaled dimensionless dynamics is obtained,

$$\frac{\partial e_\ell}{\partial \bar{ t}}=-  \frac{\partial \bar{F}(e_\ell)}{\partial e_\ell} .~~~ (8.1)$$

The underdamped dynamics \cite{R13} could be similarly scaled to quasi-universal, dimensionless form.

\subsection{Pseudospin simulations}
Microstructures can be studied using pseudospin hamiltonians, by solving self-consistency equations from  local meanfield approximations and by Monte Carlo spin simulations \cite{R29}.

\subsubsection{Local meanfield}

The completely uniform $\vec{k} =0$ meanfield contribution to the hamiltonian is $H \sim f_L (\tau) \vec{S}^2 (\vec{k} = 0)$, as the Ginzburg and compatibility  terms vanish. Thus  as $\bar{f}_L = {\bar{\varepsilon}}^2  g_L$ changes sign at transition, we have ~~~~~~~~~
$<S^2> =0$ above the transition, and $<S^2> =1$ below the transition, faithfully reproducing  the displacive transition of the original strain variable \cite{R23}. Strain textures with ${\vec k} \neq 0$,  can be captured by {\it local} meanfield  $\sigma_\ell (\vec r) \equiv ~~<S_\ell (\vec r)>$ approximations in both coordinate and Fourier space. Each spin sees a local meanfield, and there is a subtraction for consistency of  averages $<SS> \simeq \sigma \sigma$:

$$S_\ell (\vec r) S_{\ell'} ({\vec r} ') \rightarrow S_\ell (\vec r) \sigma_{\ell'}  (\vec r ') + \sigma_\ell (\vec r) S_{\ell'} ({\vec r} ')$$
$$ - \sigma_\ell (\vec r) \sigma_{\ell'} ({\vec r} ')~~(8.2a)$$
$$S_\ell (\vec k) S_{\ell'} ({\vec k})^* \rightarrow S_\ell (\vec k) \sigma_{\ell'}  ({\vec k})^* + \sigma_\ell (\vec k) S_{\ell' }({\vec k })^*$$
$$ - \sigma_\ell (\vec k) \sigma_{\ell' }({\vec k})^* .~~(8.2b)$$
This is equivalent to substituting $S = \sigma  + \delta S$ into the hamiltonian, and linearizing in $\delta S \equiv  S - \sigma$.

The meanfield hamiltonian is then, 

$$\beta H_{MF} = \sum_{\ell, \vec r} Q_\ell (\vec r) S_\ell (\vec r)=  \sum_{\ell, \vec k} Q_\ell (\vec k)^* S_\ell (\vec k), ~~(8.3)$$
where an additive  constant $-\frac{1}{2} \sum_{\vec r} Q_\ell (\vec r) \sigma_\ell (\vec r)$ on the right is suppressed. Here $Q_\ell (\vec r) \equiv \sum_{\ell', \vec r '} Q_{0 , \ell \ell'}({\vec r} - {\vec r '}) \sigma_{\ell' }(\vec r ')$, or $Q_\ell (\vec k) \equiv \sum_{\ell'} Q_{0 , \ell \ell'}({\vec k}) \sigma_{\ell'} (\vec k)$.
The same results are obtained from a ${\vec k} \neq 0$ variational approach in the weights $\rho(\vec k)$, minimizing $Trace(H \rho) - \sum_{\vec k} \rho(\vec k) \ln \rho(\vec k)$,  that has averages such as $\sigma_\ell (\vec k)= Trace \{ \rho (\vec k) S(\vec k)\}$.

The self-consistency condition for the local, nonuniform meanfield is then

$$\sigma_\ell (\vec r) = \sum_{S_\ell} S_\ell (\vec r) e^{-\beta H_{MF}} / [ \sum_{S_\ell} e^{-\beta H_{MF}}]. ~~(8.4).$$

For the $N_{OP} = 1, N_V +1 =3$ square/rectangle case, one obtains \cite{R23}

$$\sigma (\vec r) = - 2 \sinh Q(\vec r) / [ 1 + 2 \cosh Q(\vec r) ]. ~~(8.5).$$

\subsubsection{Monte Carlo simulations} 
Metropolis algorithms for nearest-neighbor  Ising models with spin components $N_{OP} =1$ and values $N_V = 2 $, are prototypical. The Monte Carlo method can be applied to spins  of  $N_{OP} = 1, 2,3$ components, with $N_V + 1 = 3, 4,5,7$  values and powerlaw anisotropic interactions. Fast Fourier Transforms (FFT) are used to easily treat 
 the full PLA compatibility potentials; this is preferable to attempting an uncontrolled truncation of the $~ 1/R^d$ interactions to some arbitrarily far-neighbor interaction, while staying only in coordinate space. The procedure is: 
  
  (i) Flips of an input  spin configuration with an input energy $H_{input}$  are made on the lattice in coordinate space, with equal probability over {\it all}  $N_V + 1$ values at a site, yielding a trial new configuration $\{\vec{S}(\vec r)\}$.  
 
  (ii) An FFT gives the
Fourier space $\{\vec{S}(\vec k)\}.$   

 (iii) The trial energy $H_{trial}$ (diagonal in Fourier space as in (7.6b)) is then evaluated. The 
 energy difference  $\Delta H = H_{trial} - H_{input}$ between configurations in the Boltzmann-factor $e^{- \beta \Delta H}$,  determines the accepted configuration. The cycle then repeats.
 
 The $N$ sites are labelled with random numbers and arranged in an increasing-value chain, so sequentially visiting every site in the chain  means every site is  visited randomly, but  once and only once, in  1 MC sweep, that  then corresponds to 1 MC step per spin.

\section{Summary}

Generalizing a procedure due to Barsch and Krumhansl, we have shown that  ferroelastic  free energies can be scaled in
 dimensionless form, defining a quasi-universality class for all materials with the same group/subgroup structural
 transition. Whereas unscaled free energies with fitted coefficients are specific to the fitted material \cite{R7}, the scaled dimensionless free energies are in a form relevant for all materials with the same ferroelastic transition.  A simplifying approximation works to leading order in the (unscaled) spontaneous- strain magnitude at transition $\lambda$, that is  typically a few percent. To this order,  the scaled Landau free energy  minima turn out in most cases to be material-independent, depending only on
the  dimensionality $d$, number of
order parameter components $N_{OP}$; and number of low-temperature
structural variants $N_V$. The minima in order-parameter space fall on the corners of `polyhedra' inscribed in `spheres' in $N_{OP}$ dimensions, with radius the scaled order parameter magnitude, that is unity at
transition. The scaled variational free energies in terms of the local order-parameter strain components, have Landau, Ginzburg, and  St Venant  powerlaw anisotropic interactions, and can be used in relaxational or underdamped dynamic simulations.  The compatibility kernels are calculated for all transitions considered, using a constraint-substitution method.

The polyhedral arrangement of minima  in the strain variables immediately suggests a 
reduced description  with strain free energies inducing `clock-zero' models, with $N_{OP}$-dimensional discrete
pseudospin vectors  at each site, pointing to  $N_V + 1$ possible states. The discrete-variable pseudospins in local meanfield and Monte Carlo simulations can be shown \cite{R29}   to reproduce previous textures  obtained in continuous-variable dynamics\cite{R11,R12,R13,R14,R28}.

Further work could include pseudospin  clock-zero simulations of more of the transitions considered here; the addition of quenched disorder; a determination   \cite{R7,R15} of  the scaled free energies and compatibility kernels for more of the 94 possible  ferroelastic transitions \cite{R6};  scaling of experimental results for different materials to explore data clustering; and simulating complex oxide models, through couplings of strains  to other fields \cite{R17}.\\

This work was supported in part by the U.S. Department of Energy,
and by ICTP, Trieste. \\ \\ \\

\section {APPENDIX: ST VENANT COMPATIBILITY KERNELS FOR FERROELASTIC TRANSITIONS}

We derive in this Appendix, the Fourier space compatibility kernels, for increasing numbers $n$ of non-order parameter strains: $n = 1, 2$
for 2D transitions, and $n = 3, 4, 5$ for 3D transitions.  

The background for compatibility ideas includes the following. Kartha et al  \cite{R11} performed Monte Carlo simulations in the displacement-vector representation, using a square/rectangle variational free energy $F ({\vec \nabla}  {\vec u})$ that, in addition to a Landau term nonlinear in the order parameters (OP), also included harmonic terms in compression and shear
$\sim (\nabla_x  u_x  + \nabla_y  u_y)^2$ and $ \sim (\nabla_y u_x  + \nabla_x u_y)^2$. They    found diagonal domain walls.  To understand the orientation,  they followed Baus and Lovett \cite{R10} and went over to the strain representation,  analytically minimizing these non-OP $ \sim {e_1}^2, {e_3}^2$ terms subject to the St Venant constraint using Lagrange multipliers. The resultant square/ rectangle compatibility kernel  explained the $\pi/ 4$ orientation preference.  However,  strain-representation simulations in the OP $e_2$, for a free energy $F( e_2)$ that explicitly included the St Venant term, were not pursued. This changeover to the OP-strain  working variable  was done in relaxational  simulations \cite{R12}, yielding the same diagonal domains. Strain-variable simulations using  compatibility kernels for other 2D cases \cite{R13, R15}, and the 3D cubic/tetragonal case \cite{R14}, found  textures as obtained  in the displacement and phase-field representations \cite{R28};  and an underdamped strain dynamics including compatibility potentials was proposed, including Langevin dynamics noise terms with powerlaw spatial correlations \cite{R13}.

Here we pursue this strain-representation project, contributing to a catalog  of 3D ferroelastic $N_{OP} \times N_{OP}$ matrix  kernels  that incorporate the three compatibility constraints (2D kernels for ferroelastic transitions were previously given in \cite{R15}). Instead of using, and solving for, three Lagrange multipliers,  we directly solve the constraints in Fourier space for three non-OP strains, and substitute in the harmonic non-OP free energies, followed by an unconstrained minimization  in any remaining non-OP strains \cite{R17}.  
The four 3D  transitions have three distinct kernels, that  are plotted in Fig 4. As a check, we apply the direct substitution  method to  find 2D kernels, previously obtained by Lagrange multipliers \cite{R11,R12,R13,R14,R15}. In Fourier space  derivatives go as  eg ${\nabla_\mu}^2  \rightarrow - {k_\mu}^2$, while on a grid with difference operators ${\Delta_\mu}^2 \rightarrow -{K_\mu }^2$ where $K_\mu \equiv 2 \sin (k_\mu /2)$. Here in compatibility equations we write for simplicity just the wave vectors like $k_\mu$, with the understanding that they can  be replaced by $K_\mu$ in grid simulations.

The compatibility constraint in minimizing  the harmonic non-OP terms $\bar{f}_{non} = \sum_i (A_i /2) |e_i (\vec k)|^2$ only affects {\it nonuniform} strains  with nonzero wave-vector. The uniform or zero wavevector parabolic terms are freely minimized by zero values, $e_i (\vec k = 0) = 0$. Thus  a prefactor
of $\nu (\vec k) \equiv 1 - \delta_{{\vec k} , 0}$ must be inserted in the results below for the kernels. The resultant sign-varying compatibility potentials  have zero spatial average             $< U(\vec R)> \sim U({\vec k} = 0) \sim \nu ({\vec k} =0) =0$, rather than a divergence  as for isotropic potentials $<U (R) > \sim \sum_{\vec R} 1/ R^d \sim \ln N$. \\ 
 
\subsection {2D transitions}

In 2D, the Fourier constraint of (2.3b)  in terms of Cartesian distortions is 
$ 2 k_x  k_y e_{xy}  - {k_y}^2 e_{x x} - {k_x}^2 e_{yy}= 0$. In terms of physical strains of (2.5) and with only one shear $s =3$, this becomes the single compatibility constraint as in (2.7),

$$O_1 e_1 + O_2 e_2 + O_3 e_3 = 0.~~ (A1)$$

For the square lattice,  the compatibility coefficients are
 $O_1 (\vec k) =- \frac{1}{\sqrt{2}} {\vec{k}}^2 ,O_2(\vec k) =+ \frac{1}{\sqrt{2}} ({k_x}^2 - {k_y}^2),O_3(\vec k) = 2k_x k_y$, while for the triangular case,$O_1 (\vec k) =-  {\vec{k}}^2 ,O_2(\vec k) = ({k_x}^2 - {k_y}^2),O_3(\vec k) = 2k_x k_y$.  In the displacement representation, for a 2D square lattice there are two independent variables $(u_x , u_y)$ per lattice point (or per unit cell). In the equivalent (symmetric) strain representation, there are three physical strains $e_1 , e_2 , e_3$ and one constraint, so there are again $3 -1 =2$ independent variables. We now derive compatibility kernels for transitions with increasing numbers of non-OP strains, $n = 1, 2$.

{\bf 1.   n = 1 cases:}

{\it Triangle/center rectangle, square/oblique, and triangle/oblique transitions:}  

For these three transitions, the two-component OP strain is $(e_2, e_3)$. The  single  non-OP strain is $e_1$, and  the harmonic term  is $\bar{f}_{non} = A_1 |e_1| ^2$. Substituting from compatibility
 $e_1 = -\sum_{\ell = 2,3} O_\ell e_\ell / O_1$ , immediately yields the $2 \times 2$ matrix kernel of  components \cite{R13}, with prefactor $\nu (\vec k)$ inserted, namely                             $U_{\ell \ell'} = \nu O_{\ell} O_{\ell'} / {O_1}^2$ so

$$U_{22} = \nu {O_2}^2 / {O_1}^2, 
~U_{33} =\nu  {O_3}^2 / {O_1}^2$$
$$ ~U_{23} = \nu {O_2 O_3} / {O_1}^2  =  U_{32}. ~~(A2a)$$ 
Or explicitly,
$$U_{22} = \nu ({k_x}^2 - {k_y}^2)^2 / { k^4}, 
~U_{33} =\nu  (2 k_x k_y)^2 / {k^4}$$
$$ ~U_{23} = \nu {2 k_x k_y} ({k _x}^2 -{k_y}^2)/{ k^4}=  U_{32}. ~~(A2b)$$ 

 Although these  2D transitions all have  $N_{OP} =2$ and the same compatibility kernel, they of course differ in their Landau or
 $g_L$ coefficients, and in the  different $N_V $  values of their nonzero pseudospin vectors (pointing to corners of a triangle, square, and hexagon, respectively).\\

{\bf 2.   n = 2 cases:}

{\it  (a)  Square/rectangle transition: }

The single OP is the deviatoric strain $e_2$.
The non-OP strains  are the 2D  compression and shear $e_1$ and $e_3$, and  the harmonic non-OP free energy is 
 $\bar{f}_{non} = \sum_{i = 1, 3} (A_i /2) | e_i (\vec k)|^2$. The compatibility condition yields $e_3 = - \sum_{\alpha = 1,2} \bar{O}_\alpha e_\alpha$, where $\bar{O}_\alpha \equiv O_\alpha / O_3$. Substituting
yields $\bar{f}_{non} = (A_1 /2) [ | e_1 (\vec k)|^2 + \sum_{  \alpha, \beta = 1,2} G_{ \alpha, \beta} e_\alpha {e_\beta}^{* }]$, where
 $G_{\alpha, \beta} \equiv (A_3 / A_1) \bar{O}_\alpha \bar{O}_\beta$.  
 
 Freely minimizing in the remaining non-OP variable  $e_1$  as $\partial f_{non}/\partial e_1 ^* (\vec k) = 0$, yields 
 $e_1 (\vec{k})=B_{12}(\vec{k})e_2(\vec{k})$.  Substituting back, 
the local non-OP term $\bar{f}_{non}$ becomes a nonlocal compatibility potential for
the OP distortions $\bar{f}_{compat} (e_2)= (A_1 / 2) U(\vec k) |e _2 (\vec k)|^2$, where
the kernel is

$$U(\vec k) = ( G_{2 2}  +  R_{22,11} ) / [ 1 + G_{11} ], ~~(A3)$$
where $R_{22,11} \equiv G_{2 2} G_{1 1} - {G_{1 2}}^2$. 
The structure is similar to  the 3D cases below. However, for the 2D case,  the remainder term $R_{22,11} =0$ so this becomes
$$A_1 U(\vec{k}) = \nu {O_2}^2 / [\{{O_1}^2 /A_1\}+\{{O_3}^2/A_3\}] .~~(A4a)$$
Or explicitly,

$$A_1 U(\vec k) = \nu  A_1( {k_x}^2  - {k_y}^2)^2 / [ k^4 + (8 A_1 /A_3)( k_x k_y)^2 ].~(A4b)$$
We  fix $2 A_1/ A_3 \simeq 1$ for simplicity, so the strength of the compatibility potential is determined by $A_1$, that is  essentially the elastic anisotropy parameter, \cite{R26} by (2.18d).
Plots of $U$ have been given elsewhere \cite{R11,R12}.
The same result is obtained through Lagrange multipliers: minimizing $\{f_{non} - \Lambda ( \sum_{\alpha =1, 2, 3} O_\alpha e_\alpha )\}$ in $e_1, e_3$
yields $e_i = \Lambda  O_i /A_i$ for $i = 1, 3$.  Demanding compatibility fixes the Lagrange multiplier 
$\Lambda = -O_2 e_2 /[ \{O_1 ^2 /A_1\} + \{O_3 ^2 /A_3\}]$, so $e_1 , e_3$ are   
in terms of  $e_2$, yielding the same kernel (A4).

As mentioned in the text, the square/rhombus is not distinct from the square/rectangle transition. Nonetheless, the OP is now $e_3$ and the non-OP are $e_1 , e_2$. The kernel is just a $2 \leftrightarrow 3$ label interchange,

$$A_1U(\vec k) = \nu {O_3}^2 /(\{{O_1}^2 /A_1\}+\{{O_2}^2/A_2\}) .~~(A5)$$

{\it (b) Rectangle/oblique}: 

The OP is again $e_3$ but the non-OP energy $\bar{f}_{non} = \sum_{\pm} (A_{\pm}/2) |e_{\pm}|^2$ is now harmonic in the combinations  $e_{+} = \frac{1}{2} (e_1 + e_2)$ and $e_{-}=\frac{1}{2} ( e_1 - e_2)$. The substitution for $e_1$ through compatibility,  and free minimization in $e_2$ yields the kernel with $O_{\pm} \equiv O_{1} \pm  O_{2}$ as
$$A_1 U(\vec k) = \nu  {O_3}^2 / [\{ {O_{+}}^2 /A_{+}\}+\{{O_{-} }^2/A_{-}\}].~~(A6)$$

\subsection{3D transitions}

We use the 3D compatibility constraints to obtain St Venant kernels for $n = 3, 4, 5$.

The 3D compatibility conditions of (2.3) in Fourier space for Cartesian distortions $Inc({\bf e}) = {\vec k}\times {\bf e}(\vec k) \times {\vec k} =0$, can be written as three equations from  diagonal components of Incompatibility,
 like $Inc(e)_{x x}=0$,
 
$$ 2k_y k_z e_{y z} =k_y ^2 e_{z z} +  k_z ^2 e_{y y} ;$$
$$ 2k_z k_x e_{z x} =   k_z ^2 e_{x x} + k_x ^2 e_{z z}  ; $$
$$ 2k_x  k_y e_{ x y} =  k_x ^2 e_{y y} + k_y ^2 e_{x x} .~~~~~(A7)$$
There are also three equations from the off-diagonal components, like $Inc(e)_{yz} = 0$,

$$ k_y k_z e_{x x} = - k_x ^2 e_{y z} +  k_x k_y e_{z x} + k_z k_x e_{x y} ;$$
$$ k_z k_x e_{y y} = - k_y ^2 e_{z x} + k_y k_z e_{x y} +  k_x k_y e_{y z} ;$$
$$k_x k_y e_{z z} = - k_z ^2 e_{x y} + k_z k_x e_{y z } + k_y k_z e_{z x}, ~~~~(A8)$$
but these are not new constraints: solving (A7) for shears and substituting,  (A8) is satisfied as an identity. In the displacement representation, for a 3D cubic lattice there are three independent variables $(u_x , u_y, u_z)$ per lattice point (or per unit cell). In the equivalent (symmetric) strain representation, there are six physical strains $e_1,..,e_6$ and three constraints, so there are again $6 -3 =3$ independent variables.

The 3D St Venant constraints  of (A7) in terms of the Cartesian distortions can be written in terms of the cubic-lattice physical distortions (2.6a), by inverting the coefficient matrix, to get

$$e_{xx} = \frac{1}{\sqrt{3}} e_1 +  \frac{1}{\sqrt{2}} e_2 + \frac{1}{\sqrt{6}} e_3 ;$$
$$e_{y y} =      \frac{1}{\sqrt{3}} e_1 -  \frac{1}{\sqrt{2}} e_2 + \frac{1}{\sqrt{6}} e_3 ;$$
$$e_{z z} =\frac{1}{\sqrt{3}} e_1 -  \frac{2}{\sqrt{6}} e_3 .~~~~(A9)$$
This yields  the 3D St Venant constraints in terms of physical distortions,

$$O^{(s)}_1 e_1 + O^{(s)}_2 e_2 + O^{(s)}_3 e_3 + O^{(s)}_s e_s = 0 ,~~(A10)$$
labelled by the three shears, $s = 4, 5, 6$. Defining 

$$\bar{O}^{(s)}_\alpha \equiv O^{(s)}_\alpha / O_s, ~~(A11)$$
 the shears $e_4, e_5, e_6$ are related to the non-shears $e_1, e_2, e_3$ by
$$e_s = - \sum_{\alpha = 1,2,3}  \bar{O}^{(s)}_\alpha e_\alpha .~~~~~(A12)$$

The cubic-lattice compatibility coefficients  $O^{(s)}_{\alpha}$ of (A10)  are evaluated from (A7) and (A9) as

 $$O^{(4)}_1= \frac{-1}{\sqrt{3}}(k_y ^2 + k_z ^2), ~O^{(4)}_2 = \frac{1}{\sqrt{2}} k_z ^2 ;$$
 $$O^{(4)}_3 = \frac{1}{\sqrt{6}} (2 k_y ^2 - k_z ^2),~ O^{(4)}_4 \equiv O_4 = k_y k_z .~~(A13a)$$ 
 
 $$O^{(5)}_1= \frac{-1}{\sqrt{3}}(k_z ^2 + k_x ^2), ~O^{(5)}_2 = \frac{-1}{\sqrt{2}} k_z ^2 ;$$
 $$ O^{(5)}_3 =  \frac{1}{\sqrt{6}} (2 k_x ^2 - k_z ^2),~ O^{(5)}_5 \equiv O_5 = k_z  k_x .~~~(A13b)$$ 

 $$O^{(6)}_1= \frac{-1}{\sqrt{3}}(k_x ^2 + k_y ^2), ~O^{(6)}_2 =   \frac{1}{\sqrt{2}}( k_x ^2 - k_y ^2) ;~$$
 $$ O^{(6)}_3 = \frac{-1}{\sqrt{6}}(k_x ^2 + k_y ^2),~  O^{(6)}_6 \equiv O_6 =k_x  k_y . ~~(A13c)$$ 
 These will be used in the cubic/tetragonal, cubic/orthorhombic, and cubic/trigonal transitions, below.

For the tetragonal-lattice physical distortions of (2.6b), the Cartesian components can be written as 
$$e_{xx} = \frac{1}{\sqrt{2}}( e_1 + e_2);~e_{y y} =\frac{1}{\sqrt{2}} (e_1 - e_2);~e_{z z} = e_3 ,~(A14)$$
and the tetragonal-lattice compatibility coefficients  from (A7) and (A14) are

 $$O^{(4)}_1= \frac{-k_z ^2}{\sqrt 2}, ~O^{(4)}_2 = \frac{k_z ^2}{\sqrt 2}, $$
 $$ ~O^{(4)}_3 =- k_y ^2 ,~ O^{(4)}_4 \equiv O_4 = k_y k_z .~~(A15a)$$ 
 
 $$O^{(5)}_1= \frac{-k_z ^2}{\sqrt 2}, ~O^{(5)}_2= \frac{-k_z ^2}{\sqrt 2} ,$$
 $$ ~O^{(5)}_3 =- k_x ^2 ,~ O^{(5)}_5 \equiv O_5 = k_z k_x .~~(A15b)$$ 
  
 $$O^{(6)}_1= \frac{-1}{\sqrt 2 }(k_x ^2 + k_y ^2), ~O^{(6)}_2 =   \frac{-1}{\sqrt 2}( k_x ^2 - k_y ^2) ,$$
 $$~ O^{(6)}_3 = 0,~~~~~~  O^{(6)}_6 \equiv  O_6 =k_x  k_y .~ ~~(A15c)$$ 
These will be used for the tetragonal/orthorhombic transition, below.

  It is useful to define a dimensionless variable, analogous to the 2D version above, namely
 
 $$G_{\alpha \beta} \equiv \sum_{s= 4,5,6} (A_s /A_1) {\bar{O}^{(s)}}_{\alpha} {\bar{O}^{(s)}}_{\beta} ,(A16a)$$ 
 where $G_{\alpha \beta} = G_{\beta \alpha}$ is symmetric, and a remainder term  as before is defined through products,
 
 $$R_{\alpha \beta, \gamma \delta} = G_{\alpha \beta} G_{\gamma \delta} - G_{\alpha \gamma} G_{\beta \delta} .~ (A16b)$$

 Then the kernels for the four  3D transitions are obtained in terms of the $G_{\alpha \beta}$, by a substitution/minimization method similar to  2D.\\ \\

{\bf  3. n = 3 case: } 

{\it Cubic/trigonal transition: }

For this transition, the OP strains  are the shears $\{e_\ell\} = \{e_s\} = e_4 , e_5 , e_6$. The non-OP  strains are $e_1, e_2, e_3$, and the harmonic term is  $\bar{f}_{non} = \sum_{i = 1,2,3} (A_i /2) |e_i (\vec k)|^2$, with deviatoric coefficients $A_2 = A_3$ by symmetry \cite{R16}.
Compatibility here gives the shear OP $e_4, e_5, e_6$ in terms of the non-OP as in (A12), $e_\ell = -\sum_{i = 1, 2, 3} {\bar{O}_{ i}}^{(s)} e_i$
or in matrix form,

\[
\left(
\begin{array}{c}
e_4 \\  
e_5 \\            
e_6   
\end{array} \right)
 = - \left(\begin{array}{ccc}
\bar{O}^{(4)}_1 & \bar{O}^{(4)}_2 & \bar{O}^{(4)}_3 \\
\bar{O}^{(5)}_1 & \bar{O}^{(5)}_2 & \bar{O}^{(5)}_3 \\
\bar{O}^{(6)}_1 & \bar{O}^{(6)}_2 & \bar{O}^{(6)}_3 
\end{array} \right )
\left(
\begin{array}{c} 
e_1 \\   
e_2 \\
e_3   
\end{array} \right). \ \  (A17) 
\]

The non-OP can be written in  terms of the OP shears by inverting the $3 \times 3$ coefficient-matrix $M$ above
to directly yield $e_i = \sum_{s = 4, 5, 6} B_{i s} e_s$. Here $B \equiv M^{-1} = (2/ Det M) N$,
where $ N \equiv   adj(M)/2$. The determinant  can be written as $Det M /2  =[ \bar{O}^{(4)}_1 N_{1 4} +  \bar{O}^{(5)}_1 N_{1 5} + \bar{O}^{(6)}_1 N_{1 6}]$. The components $N_{i s}$  can be evaluated, such as

$$N_{1 4} = -\frac{1}{2} [ \bar{O}^{(5)}_2 \bar{O}^{(6)}_3 - \bar{O}^{(6)}_2 \bar{O}^{(5)}_3 ]$$
$$ = \frac{1}{2 \sqrt{3} O_4} (k_x ^2 - k_y ^2 - k_z ^2  ). ~ (A18)$$

The elements $N_{1 4}, N_{1 5}, N_{1 6};  ... N_{3 6}$ in matrix form are:

$$
\left(
\begin{array}{ccc}
\frac{(k_x ^2 - k_y ^2 - k_z ^2  )}{2 \sqrt{3} O_4}  & \frac{(k_y ^2 - k_z ^2 - k_x ^2  )}{2 \sqrt{3} O_5} & \frac{(k_z^2 -k_x ^2 - k_y ^2)}{2 \sqrt{3} O_6} \\
\frac{(k_x ^2 + k_y ^2)}{2 \sqrt{2} O_4} & -\frac{(k_x ^2 + k_y ^2)}{2 \sqrt{2} O_5} &  - \frac{(k_x ^2 - k_y ^2 )}{2 \sqrt{2} O_6} \\
\frac{(2 k_z ^2 +k_x ^2 - k_y ^2) }{2 \sqrt{6} O_4} & \frac{(2 k_z ^2 - k_x ^2 + k_y ^2)}{2 \sqrt{6} O_5} & -\frac{(2 k_z ^2 +k_x ^2 + k_y ^2) }
{2 \sqrt{6} O_6} 
\end{array} \right ). \ \ (A19)
$$

 Evaluation  of $Det M$ with 
(A19)  yields the simple result $2 / Det M = +1$, so  $B_{i s} = N_{i s}$, and so the non-OP in terms of the OP strains are $e_i = \sum_{s = 4, 5, 6} N_{i s} e_s$ . 

The  St Venant  compatibility terms are obtained  by  simple substitution into the non-OP harmonic terms,
$\bar{f}_{non} = \sum_{i = 1,2,3} (A_i /2) |e_i (\vec k)|^2 = \sum_{ \ell, \ell'  = 4,5,6} (A_1 /2) U_{\ell \ell'} (\vec k) e_\ell (\vec k) e_{\ell'} (\vec k)^*$,
where the $3 \times 3$  compatibility matrix kernel for the cubic/trigonal transition is

$$A_1 U_{\ell \ell'} (\vec k) = \nu \sum_{i = 1,2,3} A_i N_{i \ell} N_{i \ell'} ,~~(A20)$$
and can be numerically evaluated in simulations. \\

 {\bf 4. n = 4 case:} 
  
 {\it Cubic/ tetragonal and cubic/orthorhombic transitions: }
 
 In both cases, the two-component OP are the two $3D$ deviatoric strains $(e_3, e_2)$.
 The non OP are the remaining compression and shear  strains $e_1, e_4, e_5, e_6$, and their harmonic terms are 
 $\bar{f}_{non} =(A_1 /2) |e_1|^2 +  \sum_{s =4,5,6} (A_s /2) |e_s (\vec k)|^2$, with shear coefficients $A_4= A_5 = A_6$ by symmetry \cite{R14,R16}.  From the compatibility equations (A12),
we have $e_s = - \sum_{\alpha = 1,2,3} \bar{O}^{(s)}_\alpha e_\alpha$ and hence, using the definition (A16a),

 $$\bar{f} _{non} =(A_1 /2) [ ~~|e_1|^2 + \sum_{\alpha, \beta = 1,2,3} G_{\alpha, \beta} e_\alpha {e_{\beta}}^*] .~~(A21)$$
 
Minimizing,  we get $e_1= \sum_{\ell = 2,3} B_{1 \ell} e_\ell$ with $ B_{1 \ell} = - G_{1 \ell} / ( 1 + G_{11})$
similar in structure to  the square/rectangle case. It is easy to check using (A16a) and (A13), that $G_{11}$  depends
 only on cubic-invariant combinations of $k_x , k_y , k_z$.

Substituting back into (A21), $f_{non} (e_1) = f_{compat}(e_3, e_2)$ where the St Venant term is 

$$\bar{f}_{compat}  = \sum_{\ell, \ell ' =2, 3} (A_1/2) U_{\ell \ell'} (\vec k) e_\ell (\vec k) e_{\ell'} (\vec k)^* . ~~(A22)$$
Here,  the compatibility kernel for the cubic/tetragonal and the cubic/orthorhombic transitions, is the  $2 \times 2$ matrix 

  $$ U_{\ell \ell'} =\nu  [G_{\ell \ell'} + R_{\ell \ell' , 11}]/ (1 + G_{1 1}) ,~~(A23)$$
with $R_{ \ell \ell' , 1 1} \equiv G_{\ell \ell'} G_{1 1} - G_{\ell 1} G_{\ell' 1}$. This kernel  can be numerically evaluated in simulations. \\ 
 
  As in all cases, the same results can be obtained by minimizing 
  $\{\bar{f}_{non} - \sum_{s = 4, 5, 6} \Lambda^{(s)}  [\sum_\alpha O^{(s)}_\alpha  e_\alpha]\}$ where  there are three Lagrange multipliers $\Lambda^{(s)}$. Then one finds $e_s = \Lambda^{(s)}  O_s /A_s; ~ e_1 = \sum_s \Lambda^{(s)} O^{(s)} _1 /A_1 $.  From compatibility,
  $\Lambda^{(s)} = -(A_s / O_s)\sum_{\ell = 2,3} [ \bar{O} ^{(s)}_\ell (1 + G_{11}) - \bar{O} ^{(s)}_1 G_{1 \ell}] e_\ell/ (1 + G_{11}) $, yielding the same kernel as before. This kernel was used earlier \cite{R13}, but is here and in \cite{R29} explicitly stated. \\
  
{\bf 5. n = 5 case:}

{\it  Tetragonal/orthorhombic transition:  }

For this transition, the single-component OP is one of the deviatoric strains, $e_2$. The  non-OP strains are $e_1, e_3, e_4, e_5, e_6$, and the harmonic term  is 
$\bar{f} _{non} =( A_1 /2) |e_1|^2 +  ( A_3 /2) |e_3|^2 +\sum_{s = 4, 5, 6} ( A_s /2) |e_s|^2$, with $A_5 = A_6$ by symmetry \cite{R16}. 
Substituting with (A12),  but now with the compatibility coefficients (A15) for the tetragonal case,

$$\bar{f} _{non} =\frac{ A_1}{2}[ ~\sum_{i = 1,3} (A_i / A_1) |e_i |^2 + \sum_{\alpha, \beta= 1,2, 3} G_{ \alpha, \beta} e_\alpha {e_\beta}^{* } ] ,~~(A24)$$
where the $G_{\alpha \beta}$  is defined in (A16a). 

Minimizing freely in $e_1, e_3$ and inverting a $2 \times 2$ matrix yields

$$e_1 = - [ (A_3 /A_1)G_{12} + R_{33,12}] e_2 / G_0;$$
$$~ e_3 = -[ G_{32} + R_{11,32}] e_2  / G_0,~~(A25a)$$
where $R_{\alpha \alpha, \beta \gamma}$ is defined in (A16b), and 
$$ G_0 \equiv [ (\{A_3 / A_1\} + G_{33}) ( 1 + G_{11}) - G_{13} ^2 ]. ~~(A25b)$$

The  kernel for the tetragonal/orthorhombic transition is

$$A_1 U(\vec k) = \nu [(A_3 /A_1) G_{22} + T_{2}] / G_0 ,~~~(A26).$$
with $T_{2} = (A_3 / A_1) R_{22,11} + R_{ 22,33} + \{G_{22} R_{33,11} - G_{12}R_{33,12}  - G_{32} R_{11, 32} \}$. The  kernel can be evaluated numerically in simulations.

  As a check, we take uniformity in the $z$ direction, or $k_z \rightarrow 0$, when $O_4, O_5 \rightarrow 0$, and it is clear from compatibility that $e_3 = e_4 = e_5 =0$. Then one recovers, with $4A_6 \rightarrow A_3$,  precisely the form of the square/rectangle kernel of (A4).
 
  Finally, the compatibility kernels in Fourier space  are plotted for the four 3D transitions in Fig. 4 below, reflecting the high temperature unit-cell symmetries.

\newpage

\begin{figure}
\caption{ {\it Schematic} figures of five distinct ferroelastic transitions driven by strains in 2D, with lower symmetry variants on the right: 
square to rectangle (SR); rectangle to oblique (RO); triangle to centered-rectangle (TR); square to oblique (SO), triangle to oblique (TO). For the TR case, we mark one of three equivalent lattice-point boxes, that becomes the new centred-rectangle unit cell of the first variant on the right, under the simplest distortion order-parameter. The other two equivalent  variants have similar unit-cells, that are simply from $\pm 2 \pi/3$  order-parameter rotations of that distortion. For the TO case, the equivalent rotations  are integer multiples of $2 \pi/6$.  }
\end{figure}

\begin{figure}
\caption{{\it Schematic} figures of four  ferroelastic transitions in 3D,  with the lower symmetry 
variants on the right: (a) tetragonal to orthorhombic; (b) cubic to tetragonal; (c) cubic to trigonal with two other variants not shown;
(d) cubic to  orthorhombic with four other variants not shown.}
\end{figure}

\begin{figure}
\caption{ Scaled free energy versus $N_{OP}$  order parameter  components, with minima at austenite zero state
and  $N_V$ martensite variants, and with
arrows to minima denoting pseudospin vectors.  (a) Minima on  a line for $N_{OP}=1$, $N_V=2$, square to 
rectangle  (and also tetragonal to orthorhombic,  rectangle to oblique); (b ) Minima on a triangle for $N_{OP}=2$,
$N_V=3$,  cubic to tetragonal (and  triangle to centered rectangle); (c) Minima on a square for $N_{OP}=2$, $N_V=4$, 
square to oblique polygon;   (d) Minima on hexagon for $N_{OP}=2$, $N_V=6$, for cubic to orthorhombic (and triangle to oblique);
(e) Minima (schematic)  on a tetrahedron for $N_{OP}=3$, $N_V=8$, cubic to trigonal.}
\end{figure}

\begin{figure}
\caption{ Compatibility kernel components $U_{\ell, \ell'} (k_x, k_y, k_z)$ in color plots versus $(k_x, k_y, k_z)$ for  3D transitions.  (a) Tetragonal to orthorhombic case with $U(\vec k)$. The strength of the kernel is represented by
a color coding in which dark brown represents the relative positive maxima, and dark blue the minima with zero values.The  $z$ axis is vertical and the projections shown are at $k_x=0, k_y=0$ and $k_z=0$. The maxima shown appear in the plane $k_y=0$, and minima 
for $k_x=0$ and $k_z=0$.
(b) Cubic to tetragonal (and also cubic to orthorhombic case) kernels, (i) $U_{22} (\vec k)$, (ii) $U_{33} (\vec k)$, 
and (iii) $U_{23} (\vec k)$. The three components acquire positive values,  and assume clover-leaf anisotropy in the 2D planes $k_x=0,k_y=0$ and $k_z=0$.    (c) Cubic to trigonal case, with a kernel component $U_{66} (\vec k)$. In addition to the clover-leaf pattern in the plane $k_y=0$, there is a butterfly anisotropy in the plane $k_x=0$, similar to that of the
square to rectangle kernel in 2D.}
\end{figure}

\begin{table}
\caption{Generic numbers for the scaled Landau free energies, listed in order of increasing number of structural variants $N_V$. The columns are: (1) the type of transition;  (2) spatial dimensionality; (3) order-parameter  (OP) dimensionality
$N_{OP}$ or number of pseudospin components ; (4) number of free energy minima at transition or number of pseudospin-vector states, $N_V + 1$; (5) maximum-order invariant $p_{max}$  retained in the free energy; (6) number of material coefficients $N_{mat}$, and type of scaled quasi-universality (q-u) as in text; (7) condition obeyed by OP-space angular location $\phi_m$ of minima ; (8) 'polyhedron' from minima in  $N_{OP}$ dimensions. } 

\smallskip
\begin{tabular}{||l|l|l|l||l|l|l|l}
Transition&$d$&$N_{OP}$&$N_V + 1$&$p_{max}$&$N_{mat}$, q-u&angular min condition&polyhedron\\

\hline
tetrag/orthorhombic&3&1&2 +1&6&3, first&$\sin 2\phi_m = 0$&line\\
square/rectangle&2&1&2 +1&6&3, first&$\sin 2\phi_m = 0$&line\\
square/ rhombus&2&1&2 +1&6&3, first&$\sin 2\phi_m = 0$&line\\
rectangle/oblique&2&1 &2 +1&6&3, first& $\sin 2\phi_m = 0$ &line\\
cubic/tetragonal&3&2&3 +1&4&3, first&$\sin 3\phi_m = 0$&triangle\\
triangle/centred rectangle&2&2&3 + 1&4&3, first & $\sin 3\phi_m = 0$ &triangle \\
cubic/trigonal&3&3&4 +1&4&4, second&$\sin 4\phi_m = 0,\cos^2 \theta_m =\frac{1}{3}$&tetrahedron\\
square/oblique&2&2&4 +1&6&4, second&$\sin 4\phi_m = 0$&square\\
cubic/orthorhombic&3&2&6 +1&8&6, third& $\sin 6\phi_m = 0$&hexagon\\
triangle/oblique&2&2&6 +1&8&6, third&$\sin 6\phi_m = 0$ &hexagon\\
\end{tabular}
\end{table}

\end{document}